\documentclass[12pt, a4paper]{article} %
\usepackage{graphicx} %
\usepackage[utf8]{inputenc} %
\usepackage{color} %
\usepackage{listings} %
\usepackage{url} %
\usepackage{amsfonts} %
\usepackage{amsmath} 
\usepackage{amssymb} 
\usepackage{mathrsfs} %
\usepackage{latexsym} 
\usepackage{euscript} 
\usepackage{cite} 
\DeclareTextSymbol{\degre}{OT1}{23}

\newcommand \be{\begin{equation}} %
\newcommand \ee{\end{equation}} %
\newcommand \bea{\begin{eqnarray}} %
\newcommand \eea{\end{eqnarray}} %
\newcommand \ba{\begin{array}} %
\newcommand \ea{\end{array}} %

\title{
	\textsc{Can axion-like particles explain\\ the alignments of the polarisations\\ of light from quasars?
	}
}

\author{A.~Payez, J.R.~Cudell and D.~Hutsem{\'e}kers}

\date{
}

\begin{document}

	\maketitle

	\begin{abstract}
		The standard axion-like particle explanation of the observed large-scale coherent orientations of quasar polarisation vectors is ruled out by the recent measurements of vanishing of circular polarisation.
		We introduce a more general wave-packet formalism and show that, although decoherence effects between waves of different frequencies can reduce significantly the amount of circular polarisation, the axion-like particle hypothesis is disfavoured given the bandwidth with which part of the observations were performed.
		Finally, we show that a more sophisticated model of extragalactic fields does not lead to an alignment of polarisations.
	\end{abstract}

	\section{Axion-like particles in astrophysics}

		A frequent prediction of extensions of the Standard Model of particle physics is the existence of stable weakly interacting light (sub-eV) scalar or pseudoscalar particles. The `invisible' axion~\cite{Kim:1979,SVZ:1979,Zhitnitsky:1980,DFS:1981,Weinberg:1978,Wilczek:1978} is certainly the best-known candidate, so that any particle of this kind is nowadays commonly referred to as an {\emph{axion-like particle}} (ALP) ---even though it might have nothing to do with the Peccei--Quinn solution to the strong CP problem\cite{PecceiQuinn:1977}.
Usually, the smallness of the masses of these particles is related to a very-high-energy scale where there would be new physics.
Among these ALPs, one finds for instance chameleons, coming from $f(\!\!\!~R)$ theories, but also scalar and pseudoscalar particles from Kaluza-Klein theories, super strings, or other theories beyond the Standard Model, which could be testable predictions; for recent reviews, see for instance~\cite{Raffelt:2006rj,Jaeckel:2010ni} and references therein.

		The ALPs can have a coupling to photons, as in the case for the axion~\cite{Sikivie:1983ip,Raffelt:1987im}.
		As the information we get from astrophysics comes mainly from photons from distant sources, this property makes them an appealing ingredient in many astrophysical models, and their existence could be probed by astrophysical observations.  In fact, several authors have already reported different phenomena that might find a common explanation if one supposes the existence of nearly massless axion-like particles (of mass $m\le 10^{-10}$~eV, and of coupling to photons $g\sim 10^{-11}$~GeV$^{-1}$), \textit{e.g.} the transparency of the Universe to high-energy photons~\cite{DeAngelis:2007dy}, the luminosity relations for active galactic nuclei (AGN) at different wavelengths~\cite{Burrage:2009mj}, or the high-energy cosmic rays from blazars~\cite{Fairbairn:2009zi}.
\bigskip

		Another interesting observation has to do with the distribution of position angles for polarisation of visible light coming from quasars\footnote{Hereafter, `quasar' stands for `high-luminosity AGN'.}. 
		These angles indicate the direction of maximum polarisation for each source with respect to an arbitrary direction, usually the north equatorial pole. It has been reported that the distribution of these individual preferred directions in extremely large regions of the sky ($\sim$~Gpc) is not random\cite{Hutsemekers:1998, Hutsemekers:2001,Jain:2003sg, Hutsemekers:2005}.
From the latest sample available (355 quasars), global statistical tests indicate that the probability for the observed distribution to be random is between $3~10^{-5}$ and $2~10^{-3}$, depending on the test applied\cite{Hutsemekers:2005}. 
This observation is remarkable as there is {\it a priori} no reason why one should expect such correlations over cosmological distances, larger than the most extended structures presently known in the Universe.

		This analysis also indicates that the effect is not likely to be explained by local causes (influence of our galaxy, dust, etc.) and suggests that it requires something more exotic.
		One might think that it comes from an alignment of quasar axes across the Universe. It is known~\cite{Rusk:1985,Joshi:2007yf,Borguet:2007kb,Borguet:2008tn} that, for a given quasar, the direction of preferred polarisation is related to its morphology, so that a global alignment of the axes of quasars would lead to aligned polarisations. On the other hand, if one supposes that the objects themselves are aligned, the effect should be present in radio waves. However, a study\cite{Joshi:2007yf} based on a sample of 4290 objects (52 of them being part of the sample~\cite{Hutsemekers:2005}) has shown that there is \emph{no} evidence for alignments in radio waves. It thus seems that this class of explanations is disfavoured.

		On the other hand, it has been believed that these data could naturally be explained by the mixing of light with axion-like particles in background magnetic fields~\cite{Jain:2002vx,Das:2004qka,Piotrovich:2008iy,Payez:2008pm,Hutsemekers:2008iv,Payez:2009kc,Payez:2009vi,Payez:2010xb,Agarwal:2009ic}: this would generate an alignment in visible light while leaving the polarisation of radio waves unaffected, as the mixing depends on energy. In the present paper, we show that this requires a very specific choice of magnetic fields, and that in general the alignment effect cannot be explained by the mixing. Furthermore, we shall see that, according to recent data~\cite{Hutsemekers:2010fw}, the cause of this effect cannot be photon-ALP mixing, even for magnetic fields leading to an alignment.

		In the next section, we introduce our notations and recall results for the polarisation of light described by plane waves due to the mixing of axion-like particles with photons. We then discuss why this cannot explain the data for the circular polarisation of quasars and present a wave-packet treatment of the mixing in Section~\ref{sec:newcircpoldata}. 
		Our analysis shows that, despite promising phenomenological implications, even wave-packets cannot reconcile the axion-like particle hypothesis with the full quasar sample. Further checks are made in Section~\ref{sec:ruledout}, where we use different models for the magnetic field encountered by the incoming photons.

	\section{Generalities, conventions and plane-wave formalism}\label{sec:pw}

		The mixing of photons with spin-0 particles $\phi$ changes the polarisation of light because, in a background electromagnetic field, only one specific direction of polarisation feels the interaction. 

		For pseudoscalars, the interaction Lagrangian contains a term proportional to $\phi (\vec{E}\cdot\vec{B})$. In our case, where we deal with external magnetic fields $\vec{\mathcal{B}}_e$, this reduces to $\phi (\vec{\mathcal{E}_r} \cdot \vec{\mathcal{B}}_e)$, with $\vec{\mathcal{E}_r}$, the electric field of the radiation from which the polarisation is defined.
Photons will thus mix with pseudoscalars through the projection of $\vec{\mathcal{B}}_e$ on their polarisation vector.
		Things are similar for scalars\cite{Biggio:2006im}, the main difference being that it is then the perpendicular direction which will mix, as the interaction is then related to $\phi (\vec{B}^2 - \vec{E}^2)$, and the relevant term is $\phi (\vec{\mathcal{B}_r}\cdot\vec{\mathcal{B}}_e)$, with $\vec{\mathcal{B}_r}$ the magnetic field of the radiation, so that $\vec{\mathcal{E}_r}$ has to be perpendicular to $\vec{\mathcal{B}}_e$.
		We will now stick to what happens in the pseudoscalar case for the rest of the developments, bearing in mind that our results would also hold for scalars.

		Strictly speaking, as they propagate, photons will in general have three polarisations because of the interaction with the electron plasma~\cite{Anderson:1963pc}. However, in our context, the longitudinal contribution is negligible as the electron density in the intergalactic medium is tiny. Hence it is sufficient to consider the projection of the magnetic field onto a plane perpendicular to the direction of propagation, \textit{i.e.} the transverse part of $\vec{\mathcal{B}}_e$, noted $\vec{\mathcal{B}}$.
		We define an orthogonal basis made of the direction of propagation of light and of two directions $\vec{e}_{\parallel}$ and $\vec{e}_{\perp}$, parallel and perpendicular to the transverse field. Any light beam going in the $z$ direction will then be written as $\vec{\mathcal{E}}_r(z,t)=\mathcal{E}_{r_{\parallel}}(z,t)\vec{e}_{\parallel} + \mathcal{E}_{r_{\perp}}(z,t)\vec{e}_{\perp}$.
		\bigskip

		We can now express the evolution of an electromagnetic wave in an external magnetic field $\vec{\mathcal{B}}$ and obtain the evolution of its Stokes parameters, due to the mixing with pseudoscalars.
		Stokes parameters are particularly interesting because they can fully describe the polarisation of light, and because they are the quantities that observers directly measure, as they can be built out of intensities. 
		We use the following definitions:
                \bea
                \left\{
                    \begin{array}{llll}
                        I(z) &=& \langle\mathcal{I}(z,t)\rangle \ \!= \langle \mathcal{E}_{r_{\parallel}}\mathcal{E}^*_{r_{\parallel}} + \mathcal{E}_{r_{\perp}}\mathcal{E}^*_{r_{\perp}}\rangle\\
                        Q(z) &=& \langle\mathcal{Q}(z,t)\rangle = \langle \mathcal{E}_{r_{\parallel}}\mathcal{E}^*_{r_{\parallel}} - \mathcal{E}_{r_{\perp}}\mathcal{E}^*_{r_{\perp}}\rangle\\
                        U(z) &=& \langle\mathcal{U}(z,t)\rangle = \langle \mathcal{E}_{r_{\parallel}}\mathcal{E}^*_{r_{\perp}} + \mathcal{E}^*_{r_{\parallel}}\mathcal{E}_{r_{\perp}}\rangle\\
                        V(z) &=& \langle\mathcal{V}(z,t)\rangle = \langle i(- \mathcal{E}_{r_{\parallel}}\mathcal{E}^*_{r_{\perp}} + \mathcal{E}^*_{r_{\parallel}}\mathcal{E}_{r_{\perp}}) \rangle.
                    \end{array} \right. 
                    \label{eq:StokesE}
                \eea
	$U$ and $Q$ represent linear polarisation, $V$ the circular one and $I$ the intensity; these quantities are averaged over time at a given distance $z$ from the source.
	One often normalises these parameters by the intensity $I$ to enable comparisons between different sources (\textit{e.g.}, $v=\frac{V}{I}$) and, as $U$ and $Q$ depend on the choice of axes, one also introduces the linear polarisation degree and the polarisation degree, respectively
		\be
			p_{lin} = \frac{\sqrt{Q^2 + U^2}}{I}\qquad \textrm{and} \qquad p_{tot} = \frac{\sqrt{Q^2 + U^2 + V^2}}{I}.
		\label{eq:poldegree}
		\ee

	The parameters~\eqref{eq:StokesE} have a property of additivity: a radiation described by a set ($I,Q,U,V$) can always be decomposed as the sum of two other ones, described by ($I_1,Q_1,U_1,V_1$) and ($I_2,Q_2,U_2,V_2$), as long as $I_1+I_2=I$, etc. In particular, a partially polarised beam can be described by the weighted sum of a fully polarised beam and of an unpolarised one: it is thus sufficient to discuss these two cases to obtain the most general case.

	Formally, one can write a general initially fully polarised light beam of mean frequency $\omega$ and width $\Delta\omega$ as:
	\be
		\vec{\mathcal{E}}_r(z,t) = \sin(\varphi_0)E_{\parallel}(z,t;\omega;\Delta\omega)\vec{e}_{\parallel} + \cos(\varphi_0)E_{\perp}(z,t;\omega;\Delta\omega)\vec{e}_{\perp},
		\label{eq:fullypollight}
	\ee
	where $E_{\parallel}\vec{e}_{\parallel}$ and $E_{\perp}\vec{e}_{\perp}$ are fully linearly polarised beams with polarisations respectively parallel and perpendicular to $\vec{\mathcal{B}}$, and with identical intensities. Initially, $E_{\parallel}$ and $E_{\perp}$ (which may differ by a phase-shift) have the same behaviour and shape, but this will change as they propagate. The angle $\varphi_0$ gives the initial direction of $\vec{\mathcal{E}}_r$.
	On the other hand, for an initially unpolarised beam, the Stokes parameters can be thought of as an incoherent average over $\varphi_0$ of the Stokes parameters of fully polarised beams.
\bigskip

		The way the polarisation parameters evolve in a magnetic field $\vec{\mathcal{B}}$ can be derived starting from a suitable Lagrangian density taking into account the interaction. For pseudoscalars, we use:
			\begin{equation}
				 \mathcal{L} =  \frac{1}{2}\ (\partial_{\mu}\phi) (\partial^{\mu}\phi) - \frac{1}{2}\ m^2\phi^2 - \frac{1}{4}\ F_{\mu\nu} F^{\mu\nu} + \frac{1}{4}\ g \phi F_{\mu\nu}\widetilde{F}^{\mu\nu}, \label{eq:lagrangian}
			\end{equation}
			where $\widetilde{F}^{\mu\nu}\equiv\frac{1}{2}\ \epsilon^{\mu\nu\rho\sigma}F_{\rho\sigma}$ is the dual of the electromagnetic tensor, $m$ is the pseudoscalar mass and $g$ is the dimension-minus-one coupling constant of the interaction between pseudoscalars and photons.

		At this stage, one can take into account plasma effects with the inclusion of the plasma frequency $\omega_p$~\cite{Carlson:1994yqa,Deffayet:2001pc}:
		\be
			\omega_p \equiv \sqrt{\frac{4\pi\alpha n_e}{m_e}} = \sqrt{\frac{n_e}{10^{-6}\textrm{cm}^{-3}}}\ {3.7~10^{-14}\textrm{eV}},\label{eq:plasmafhz}
		\ee
which acts as an effective mass for the propagating electromagnetic field; $n_e$ is the electron number density.

		We first consider a constant magnetic field region in order to introduce the consequences of the mixing on polarisation. We use in this case a typical field strength of $0.3~\mu$G and a typical coherence scale of 10~Mpc, as in our supercluster~\cite{Vallee:1990, Vallee:2002, Giovannini:2003yn, Burrage:2008ii,Vallee:2011}, with an electron density such that $\omega_p=3.7~10^{-14}$~eV~\cite{Vallee:1990,Vallee:2011,Das:2004qka,Burrage:2008ii,Kravtsov:2002ac}.
The equations for the electromagnetic potential and the pseudoscalar field are then found to be:\footnote{
	This is obtained to lowest order in $g\mathcal{B}$.
}
	    \be
		    \Bigg[\Big(\omega^2 + \frac{\partial^2}{\partial z^2}\Big) -
	                                \left(
	                                \begin{array}{ccc}
	                                 {\omega_p}^2 & 0            & 0\\
	                                0            &  {\omega_p}^2 & - g \mathcal{B} \omega\\
	                                0            & - g \mathcal{B} \omega & m^2
	                                \end{array} \right)\Bigg] \left(\!\! \begin{array}{c}A_{\perp}(z) \\ A_{\parallel}(z) \\\phi(z)\end{array}  \!\!\right) = 0,\label{eom_planewaves}
	    \ee
for eigenstates of energy $\omega$, in the timelike axial gauge $A^0=0$, so that $\vec{E}_{\perp,\parallel} = i\omega\vec{A}_{\perp,\parallel}$, and after a rephasing of $\phi(z)$. Note that Faraday rotation is not included in the discussion as its effect is irrelevant in the range of frequencies we are interested in.

The mass matrix in Eq.~\eqref{eom_planewaves} is not diagonal, which means that  $A_{\parallel}$ and $\phi$ are not the eigenmodes of propagation inside $\vec{\mathcal{B}}$. These are found by diagonalisation and correspond to two new mass eigenvalues, $\mu_+$ and $\mu_-$, that depend on $\omega$:
	\be
			{\mu_{\pm}}^2 = \frac{1}{2}({\omega_p}^2 + m^2) \pm r_{mix},
	\label{eq:mupm}
	\ee
with
	\be
		r_{mix} \equiv \frac{1}{2}\sqrt{{{(2g\mathcal{B}\omega)}^2 + (m^2 - {\omega_p}^2)}^2},
	\ee
and the mixing angle:
	\be
		\theta_{mix} = \frac{1}{2} \textrm{atan}\left(\frac{2g\mathcal{B}\omega}{m^2 - {\omega_p}^2}\right).\label{eq:thetamix}
	\ee

	For the moment, we always take $\phi(0)=0$, \textit{i.e.} we only consider incoming photons. The evolution of the Stokes parameters of Eq.~\eqref{eq:StokesE} inside a magnetic field region for a plane-wave beam $\vec{\mathcal{E}}_r$ described initially by $I_0, Q_0, U_0$ and $V_0$ are then\footnote{The only simplification we have made to obtain Eq.~\eqref{eq:Stokes_alternative} is to suppose $\omega^2\gg{\mu_{\pm}}^2$, which indeed holds in all the applications we are interested in~---as in most astrophysical situations where the mixing takes place in faint background magnetic fields.}:
                \bea
                \left\{
                    \begin{array}{llll}
                        I(z) &=& I_0 - \frac{1}{2}\left(I_0 + Q_0\right) \sin^2 2\theta_{mix} \sin^2\left(\frac{1}{2}\frac{r_{mix}}{\omega}z\right)\\
                        Q(z) &=& I(I_0\longleftrightarrow  Q_0)\\
                        U(z) &=& U_0\big\{  {(\textrm{s}_{mix})}^2 \cos\left({\left(\textrm{c}_{mix}\right)}^2\ \frac{r_{mix}}{\omega}z\right) + {(\textrm{c}_{mix})}^2 \cos\left({\left(\textrm{s}_{mix}\right)}^2\ \frac{r_{mix}}{\omega}z\right) \big\}\\
			&& \!\!\!\!- V_0\big\{ {(\textrm{s}_{mix})}^2 \sin\left({\left(\textrm{c}_{mix}\right)}^2\ \frac{r_{mix}}{\omega}z\right) \ \!- {(\textrm{c}_{mix})}^2 \sin\left({\left(\textrm{s}_{mix}\right)}^2\ \frac{r_{mix}}{\omega}z\right) \big\} \textrm{sign}(\theta_{mix})\\
                        V(z) &=& U(U_0\rightarrow V_0, V_0\rightarrow-U_0),
                    \end{array} \right. 
                    \label{eq:Stokes_alternative}
                \eea
	with $\textrm{c}_{mix}\equiv\cos(\theta_{mix})$ and $\textrm{s}_{mix}\equiv\sin(\theta_{mix})$.
	Hence, the evolution of Stokes parameters can be expressed in such a way that all the effects of the mixing with pseudoscalars in a given $\vec{\mathcal{B}}$ depend only on two dimensionless parameters: the mixing angle $\theta_{mix}$ and the quantity $(\frac{r_{mix}}{\omega}z)$.
\bigskip

	Eq.~\eqref{eq:Stokes_alternative} implies dichroism and birefringence; see, \textit{e.g.}\cite{Payez:2008pm}. Dichroism is the selective absorption of one direction of polarisation: it modifies the linear polarisation of light. This effect is seen in the evolution of the Stokes parameter $Q(z)$ which compares the intensity in the two orthogonal directions.
	The total intensity $I(z)$ of course follows the same behaviour. The pair ($I$, $Q$) is directly sensitive to the modifications of the amplitude of photons due to on-shell pseudoscalars.

	Birefringence, on the other hand, indicates that linear and circular polarisations convert into each other. This strong connection between the two is explicit in the evolution of $U(z)$ and $V(z)$. The pair ($U$, $V$) is directly sensitive to the phase-shift induced by virtual pseudoscalars: pure $U$ requires a zero phase-shift and pure $V$ requires a $\frac{\pi}{2}$ phase-shift between $\mathcal{E}_{r,\parallel}$ and $\mathcal{E}_{r,\perp}$.

	Note that for an initially unpolarised light beam, while $I(z)$ and $Q(z)$ will evolve due to pseudoscalar-photon mixing, $U(z)$ and $V(z)$ will remain zero. The reason is clear: for unpolarised light, the concept of phase-shift does not make sense.
Similarly, if a linearly polarised beam points either exactly in the magnetic field direction or perpendicularly to it, \textit{i.e.} $Q(0)\neq0$ and $U(0)=0$, there cannot be any induced phase-shift and, therefore, no induced circular polarisation.

	\subsection{Initially unpolarised light and dichroism}
\begin{figure}[h!!]
			\begin{center}
				\includegraphics[width=0.7\textwidth]{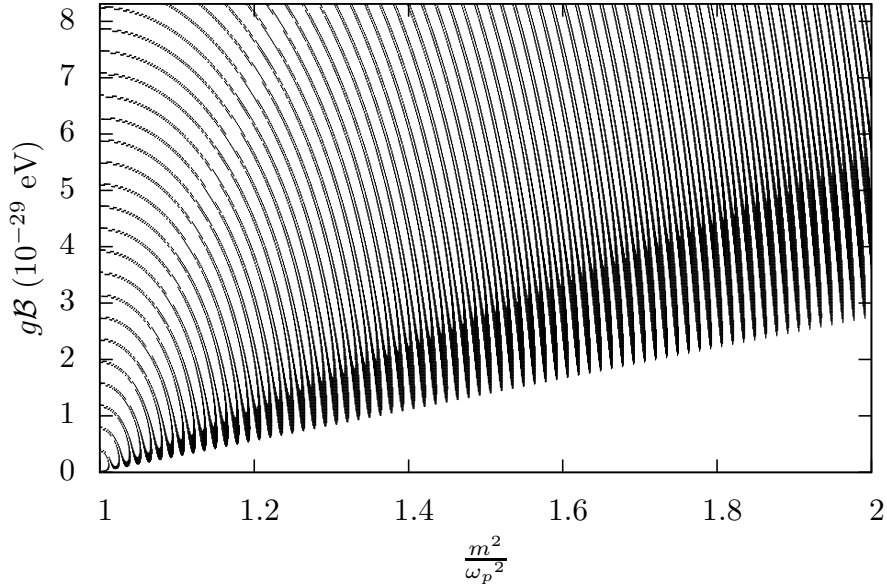}
			\end{center}

			\caption{Parameters such that the linear polarisation generated through pseudoscalar-photon mixing in a transverse magnetic field region lies between 0.5 and 2\% in the case of initially unpolarised light of wavelength $\lambda = 500$~nm. The magnetic field has been chosen to be 10~Mpc long; the plasma frequency is kept fixed at $3.7~10^{-14}$~eV.
			}
			\label{fig:p_lin_ok}
		\end{figure}

	To reproduce coherent alignments of the polarisations of light from quasars, dichroism will be the main mechanism leading to the generation of a systematic amount of linear polarisation: as estimated in~\cite{Hutsemekers:2010fw}, this has to be at least 0.5\% and certainly not more than 2\% to explain data.	
	In order to present only this additional linear polarisation, and avoid the arbitrariness of the initial one, we first consider unpolarised light beams.\footnote{Note that even for partially linearly polarised light, which for quasars is at the 1\% level~\cite{Stockman:1984qz,Berriman:1990qz,Sluse:2005}, the unpolarised contribution will remain the dominant component.} For a given travelled distance $z$, and for a fixed value of $\omega_p$, we study the space of parameters that reproduce the observed linear polarisation. The result is shown in Fig.~\ref{fig:p_lin_ok}. Note that we do not display pseudoscalar masses smaller than $\omega_p$ since the linear polarisation degree is an even function of ($m^2 - {\omega_p}^2$). This remains true as long as there is no initial circular polarisation.

	What we learn from Fig.~\ref{fig:p_lin_ok} is that as long as $m$ and $\omega_p$ are of the same order of magnitude, the mixing effect can in principle be observable and reproduce the observations, even in faint ---but extended--- magnetic fields ($\mathcal{B}=0.1~\mu$G and $g=10^{-11}$~GeV$^{-1}$ correspond to $g\mathcal{B} = 1.95~10^{-29}$~eV). The plasma frequency is very small in super-clusters, and even smaller in cosmic voids, where only upper bounds exist to date in the literature~\cite{Csaki:2001jk,Mortsell:2002dd,DeAngelis:2007yu,Mirizzi:2006zy}. With this scenario, the observations would then seem to be compatible with the existence of nearly massless axion-like particles. Note that the mixing depends on $m$ and $\omega_p$ separately and that nothing special happens if we take $\omega_p = 0$.
\bigskip

\begin{figure}[h]
			\begin{center}
					\includegraphics[width=0.8\textwidth]{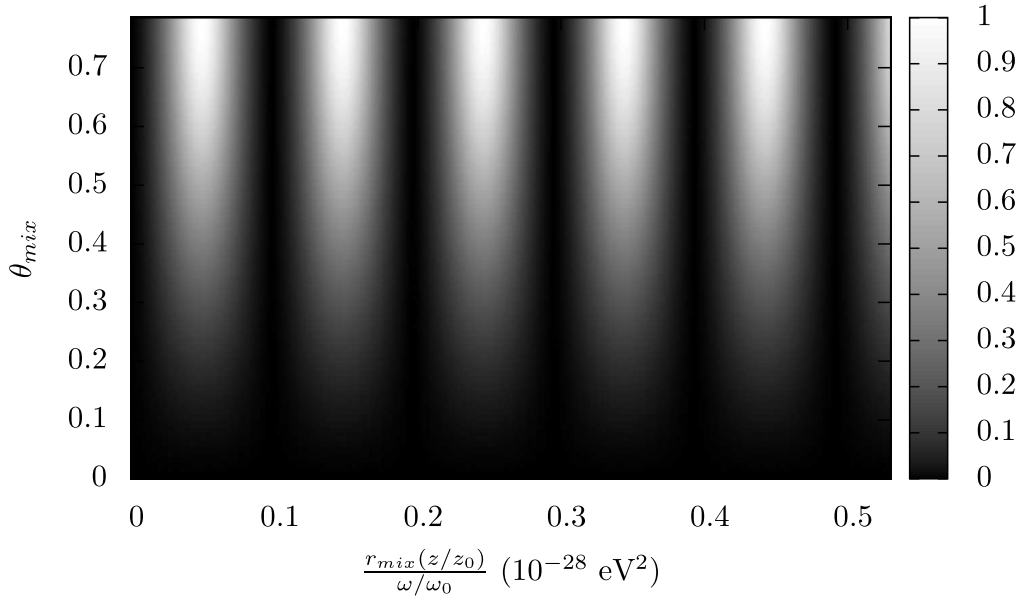}
			\end{center}
			\caption{
Linear polarisation degree (shown in the right-hand box) generated through pseudoscalar-photon mixing in a transverse magnetic field region in the case of initially unpolarised light. For convenience, we have introduced $\omega_0 = 2.5$~eV (\textit{i.e.} $\lambda_0 = \frac{2\pi}{\omega_0} =500$~nm) and $z_0=1.6~10^{30}$~eV$^{-1}$ ($\simeq10$~Mpc).
			}
			\label{fig:rtheta}
		\end{figure}

	For initially unpolarised light, the linear polarisation degree \eqref{eq:poldegree} in terms of $\frac{r_{mix}}{\omega}z$ and $\theta_{mix}$ takes the form:
	\be
		p_{lin}(z)=\frac{\frac{1}{2}\sin^2 2\theta_{mix}\sin^2[\frac{1}{2}\frac{r_{mix}}{\omega}z]}
						   {1 - \frac{1}{2}\sin^2 2\theta_{mix}\sin^2[\frac{1}{2}\frac{r_{mix}}{\omega}z]}.
	\label{eq:plinrtheta}
	\ee
	We illustrate Eq.~\eqref{eq:plinrtheta} in Fig.~\ref{fig:rtheta} ---strictly speaking $\theta_{mix}\in[-\frac{\pi}{4},\frac{\pi}{4}]$ but $p_{lin}$ is an even function of it.
	Note that the maximum linear polarisation is entirely determined by $\theta_{mix}$, while the details of the oscillatory behaviour with $z$ are independently controlled by $r_{mix}$.
This can be physically understood as $\theta_{mix}$ determines how much the particles mix, while $r_{mix}$ has to do with the difference of mass eigenstates and is thus related to the wavelength of the oscillation.
	As long as the Lagrangian~\eqref{eq:lagrangian} makes sense, the oscillatory pattern in $\frac{r_{mix}z}{\omega}$ repeats itself unchanged to infinity: all the physics can thus be studied in a small interval.

\begin{figure}[h!!]
			\begin{center}
					\includegraphics[width=0.8\textwidth]{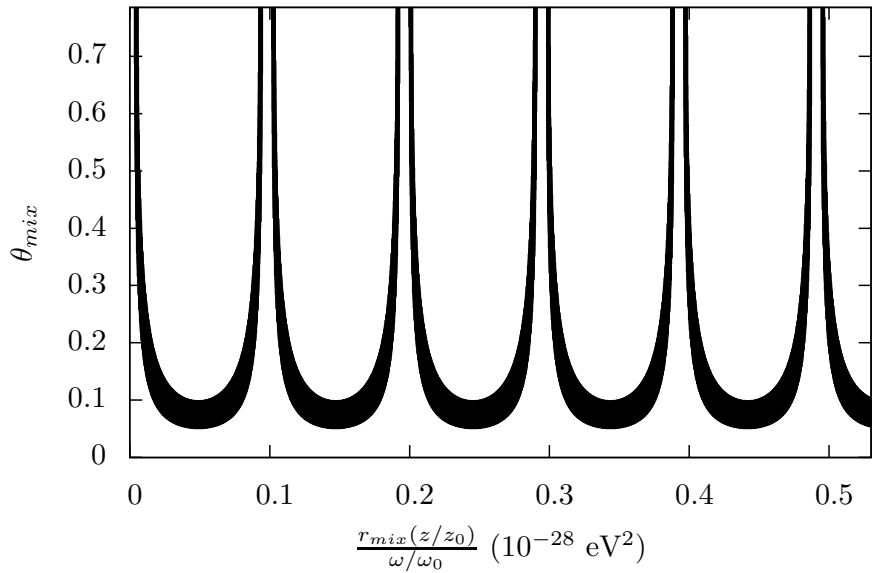}
			\end{center}
			\caption{Same as Fig.~\ref{fig:rtheta} but such that $p_{lin}\in[0.005,0.02]$. For fixed $z=z_0$ and $\omega=\omega_0$, this is equivalent to a reparametrisation of Fig.~\ref{fig:p_lin_ok}.
			}
			\label{fig:rthetaplinok}
		\end{figure}

	Now, if we are only interested in the parameters able to explain quasar linear polarisation data, we get Fig.~\ref{fig:rthetaplinok}. We can also consider the average over a period in $z$ of the additional polarisation, and impose that it lies between 0.005 and 0.02. This gives an allowed range of values for $\theta_{mix}$, which is:
	\be
		0.07\leq |\theta_{mix}| \leq0.14.
		\label{eq:thetarange}
	\ee

	\subsection{Initially polarised light and birefringence}

	For astronomical sources, processes leading to the production of circularly polarised light are rare. Generally, most quasars only emit partially linearly polarised light; their polarisation degree is typically around 1\%~\cite{Stockman:1984qz,Berriman:1990qz,Sluse:2005}. In the following, we suppose that the initial distribution of polarisation angles is random, so that the radiation can be described by random initial values for $q(0)$ and $u(0)$, with $p_{lin}(0)=\sqrt{u^2(0)+q^2(0)}=0.01$, and we assume no initial circular polarisation: $v(0)=0$.
Note that the observed linear polarisation of the quasars in the sample is also of the order of 1\%, and is believed to be mainly of intrinsic origin~\cite{Hutsemekers:1998,Hutsemekers:2001,Hutsemekers:2005,Sluse:2005,Hutsemekers:1998pp,Schmidt:1999p,Lamy:2004yz}.
	\bigskip

	Now, while the mixing can generate enough linear polarisation to reproduce the effect via axion-like particles, birefringence is expected to lead to an observable amount of circular polarisation, see \textit{e.g.}~\cite{Maiani:1986md,Raffelt:1987im,Hutsemekers:2010fw}.
	Indeed, as readily seen in Eq.~\eqref{eq:Stokes_alternative}, a linearly polarised light beam (with non-zero $u(0)$) will develop a circular polarisation as it propagates.\footnote{
This is also true for low-mass axion-like particles, even if the induced phase-shift $\Phi$ drops quickly as the mass decreases: in the weak-mixing limit \cite{Cameron:1993mr},
	\begin{equation}
		\Phi = \theta_{mix}^2 \left[ \frac{m^2 z}{2 \omega} - \sin\left(\frac{m^2 z}{2 \omega}\right) \right].
	\end{equation}
	In this astrophysical context, $\Phi$ is not small when the considered magnetic field regions are huge.
} From a technical point of view, an initial angle of $\frac{\pi}{4}$ with the direction of the external magnetic field leads to the maximal amount of generated circular polarisation; it corresponds to $u(0)=p_{lin}(0)$.

	Moreover, even for light coming from a single quasar, a number of regions with different uncorrelated magnetic fields will be encountered on the way towards us. It is thus impossible to avoid $u(0)\neq0$ at the beginning of some of these regions.

\begin{figure}
			\begin{center}
					\includegraphics[width=0.8\textwidth]{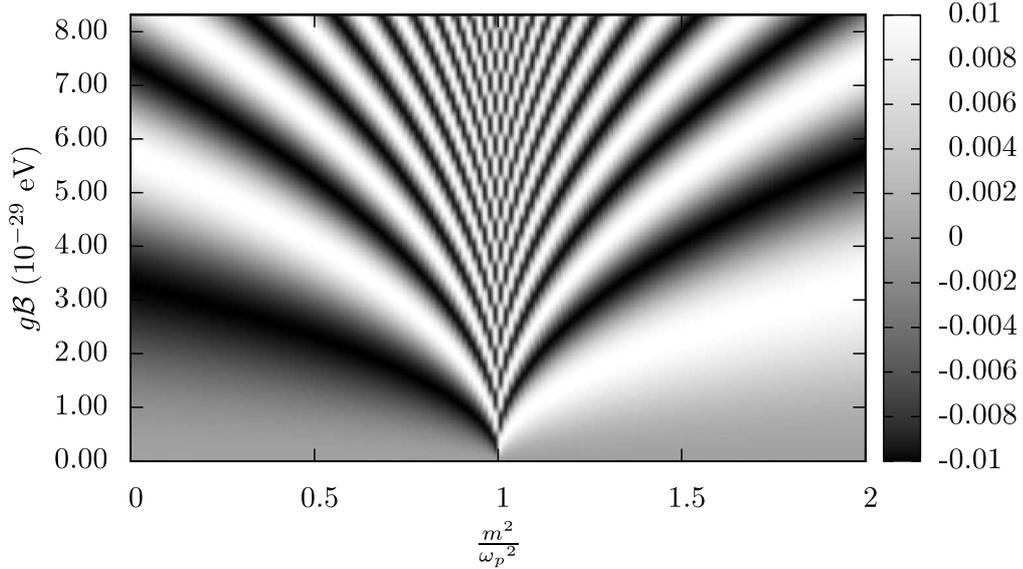}
			\end{center}
			\caption{Circular polarisation $v$ generated through pseudoscalar-photon mixing in a transverse magnetic field region, in the case of initially partially polarised light of wavelength $\lambda=500$~nm with $u(0)=0.01$. The plasma frequency and the size of the magnetic region are the same as in Fig.~\ref{fig:p_lin_ok}.
			}
			\label{fig:IPpcirc_pl}
		\end{figure}

\begin{figure}
			\begin{center}
					\includegraphics[width=0.8\textwidth]{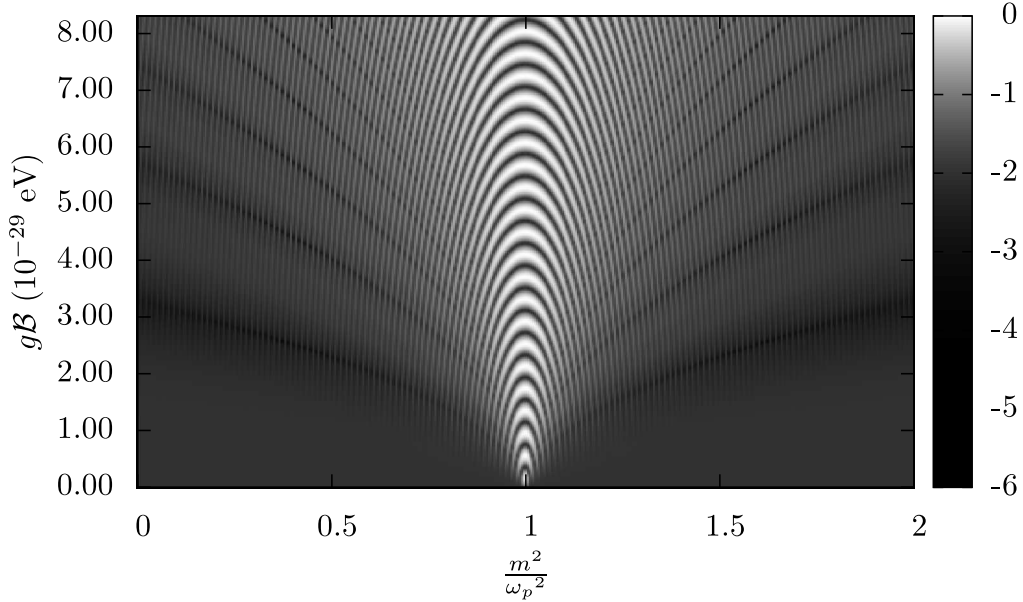}
			\end{center}
			\caption{Same as Fig.~\ref{fig:IPpcirc_pl} but for the linear polarisation degree. Note that the right-hand box gives the base-10 logarithm of the linear polarisation.
			}
			\label{fig:IPplin_pl}
		\end{figure}

	As we did for linear polarisation in the initially unpolarised case, we now calculate the circular polarisation $v$ predicted by pseudoscalar-photon mixing when light is described by plane waves with $\lambda=500$~nm, for an initial linear polarisation of $1\%$.
	We show in Fig.~\ref{fig:IPpcirc_pl} the circular polarisation generated in one magnetic region, for $u(0)=p_{lin}(0)$, and in Fig.~\ref{fig:IPplin_pl} the corresponding linear polarisation; as Stokes parameters are additive, any other case can be obtained from this one.
	Note that, as long as $v(0)=0$, the circular polarisation is an odd function of ($m^2 - {\omega_p}^2$).
Fig.~\ref{fig:IPpcirc_pl} indicates that a large region of the parameter space leads to an observable circular polarisation, \textit{i.e.} most of the quasars should be circularly polarised, with a circular polarisation of the order of the observed linear polarisation.
\bigskip

\section{New data on circular polarisation and decoherence}\label{sec:newcircpoldata}

	Recently, the circular polarisation of quasars belonging to the sample~\cite{Hutsemekers:2005} has been accurately measured in visible light. This analysis~\cite{Hutsemekers:2010fw} shows that, except for two specifically highly polarised blazars, which might be intrinsically circularly polarised, the objects have a circular polarisation consistent with zero.

\begin{figure}[h!!]
			\begin{center}
					\includegraphics[width=0.8\textwidth]{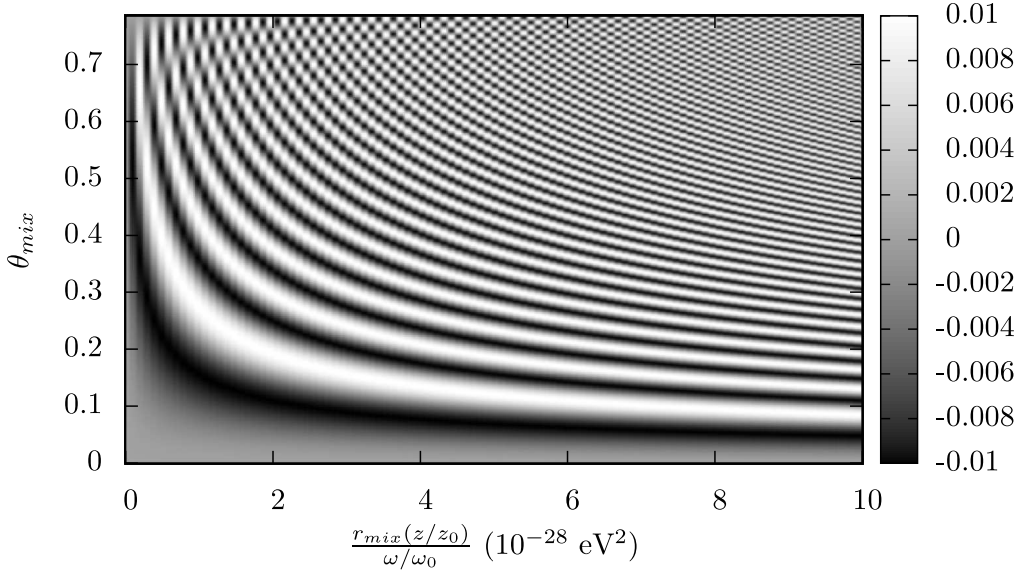}
			\end{center}
			\caption{Circular polarisation $v$ generated through pseudoscalar-photon mixing in a transverse magnetic field region in the case of initially partially polarised light with $u(0)=p_{lin}(0)=0.01$. The values of $\omega_0$ and $z_0$ are the same as the ones introduced in Fig.~\ref{fig:rtheta}.
			}
			\label{fig:rthetapcirc}
		\end{figure}

	This is clearly in contradiction with the results presented above for circular polarisation; we show this in Fig.~\ref{fig:rthetapcirc}, in terms of $\left(\frac{r_{mix}z}{\omega}\right)$ and $\theta_{mix}$.
	One sees that $v$ is of the same order of magnitude as $p_{lin}$, except in a small region (compare with Fig.~\ref{fig:rthetaplinok}): $\frac{r_{mix}(z/z_0)}{(\omega/\omega_0)}\lesssim0.2~10^{-28}$~eV$^{-2}$, and $|\theta_{mix}|$ with values similar to the ones in Eq.~\eqref{eq:thetarange}.

	Keeping an additional linear polarisation of the order of 1\% while suppressing the circular one then requires a considerable amount of fine-tuning of the masses, or smaller regions of magnetic field. The latter seems excluded as the correlations of polarisations over huge distances require magnetic fields to be coherent in large regions.
	To illustrate the need for fine-tuning, we can give a pseudoscalar mass for which enough linear polarisation is created (the maximum of $p_{lin}$ is determined by $\theta_{mix}$), and such that circular polarisation is much smaller than the linear one (by choosing a suitable $r_{mix}$).\footnote{We can consider $\theta_{mix}$ as determined in the unpolarised case: the additional polarisation will indeed not be very different here, as by requiring no $v$, we essentially constrain $u$ not to change.}
First, we write the pseudoscalar mass as a function of $\theta_{mix}$, and of $r_{mix}$, for a fixed value of $\omega_p$:
	\be
		m = \sqrt{{\omega_p}^2 + 2 r_{mix} \cos\left(2\theta_{mix}\right)},\quad\textrm{if }m > \omega_p;
	\ee
	\be
		m = \sqrt{{\omega_p}^2 - 2 r_{mix} \cos\left(2\theta_{mix}\right)},\quad\textrm{if }m < \omega_p.
	\ee
	For $z=z_0=10~$Mpc and $\omega= \omega_0=2.5$~eV, $u(0)=1\%$, and using $\theta_{mix} = 0.1$, we then obtain that the only allowed ALP masses able to reproduce data would be such that:
	\be
		\frac{m}{\omega_p} \in \left[0.99,1.01\right].\label{eq:finetuned}
	\ee
	This is a very fine-tuned situation, especially given that we have allowed $v$ to be as large as 0.1\% in this example.\footnote{If we require $v<0.01\%$, the range of allowed values for the mass shrinks to $m \in \left[0.998,1.002\right]\omega_p$.} Note also that the plasma frequency is expected to vary along the light trajectory, so that~\eqref{eq:finetuned} cannot be maintained.
	These data thus disfavour the ALP hypothesis in the plane-wave case.

\subsection{A wave-packet treatment}\label{sec:wptreatment}

We now consider the possibility of reducing circular polarisation by considering wave packets, which automatically include the possibility of decoherence between waves of different frequencies. As circular polarisation is a matter of phase-shifts, decoherence effects can significantly reduce it.
There is also a natural observational reason for taking into account this effect: astronomers perform polarimetric measurements in given ranges of frequencies, with given filters. In the case at hand, some data were obtained in white light (unfiltered), and some with the so-called `Bessell V-filter'~\cite{Sterken:1992ap,EFOSC2:2008}.

	From the Lagrangian \eqref{eq:lagrangian}, the system of relevant equations is:
	\bea
		\left\{
			\begin{array}{llll}
		(\square + {\omega_p}^2) E(z,t) -g\mathcal{B} \partial^2_t\phi(z,t) &=& 0\\
		(\square + m^2) \phi(z,t) + g\mathcal{B} E(z,t) &=& 0,
			\end{array}
			\label{eq:system}
		\right.
	\eea
	where we simplify the notation: from now on, $E\equiv E_{\parallel}$. Note that the solution for $E_{\perp}$ will simply be that for $E_{\parallel}$, with $g\mathcal{B}$ set to zero.

	We consider the case in which a wave packet is sent into a region of constant magnetic field $\mathcal{B}$, starting at $z=0$, and use wave packets in $\omega$:
	\be
		E (z,t) = \int_{-\infty}^{\infty} d\omega\    e^{- i \omega t} \widetilde{E} (z,\omega)
		\qquad\textrm{and}\qquad
		\phi (z,t) = \int_{-\infty}^{\infty} d\omega\    e^{- i \omega t} \widetilde{\phi} (z,\omega).
		\label{eq:decompsuromega}
	\ee
	Equations~\eqref{eq:system} have then to be satisfied by the integrands of \eqref{eq:decompsuromega} in each region, with $\mathcal{B}=0$ if $z\leq0$ (region I), and $\mathcal{B}\neq0$ if $z\geq0$ (region II); the solutions are given in Appendix~\ref{app:wp}. For the rest of the discussion, the incident packet in the first region has the initial shape:
	\be
		\widetilde{{E}}_{i, I}(z = 0,\omega) = e^{-\frac{a^2}{4}(\omega-\omega0)^2}.
		\label{eq:gaussianpacket}
	\ee

	\subsubsection{Size of the wave packets}

	Continuum light coming from quasars, at least in UV and visible wavelengths, is thermally emitted in the accretion disk. In order to obtain an estimate of the wave-packet size in this case, we can start with results for black-body radiation: we decompose the accretion disk into a concentric collection of black bodies of different temperatures at different radii~\cite{Shakura:1973bh,Kochanek:2007ta}.\footnote{Strictly speaking, one would then have to average the results obtained for different black bodies over the range of frequencies actually observed.} For a black-body radiation of Wien wavelength $\lambda_{w}$, estimates of the longitudinal coherence length $l_c$ have been obtained in interferometry~\cite{Donges:1998}: $l_c \simeq\lambda_{w}\simeq \bar{\lambda}$, with $\bar{\lambda}$, the mean wavelength of the radiation.
	The relation we use for the value of $a$ which enter Eq.~\eqref{eq:gaussianpacket} is then:\footnote{From~\cite{Donges:1998}, this can change slightly, depending on the definition of the full-width-at-half-maximum in frequencies.}
	\be
		a = \frac{2\sqrt{\ln(2)}}{\pi}\bar{\lambda},
	\ee
	and the initial full-width at half-maximum in position is:
	\be
		\Delta z\simeq\sqrt{2\ln(2)}a,
	\ee
	which is thus of the order of the wavelengths considered.

		\begin{figure}[h]
			\begin{center}
					\includegraphics[width=0.49\textwidth]{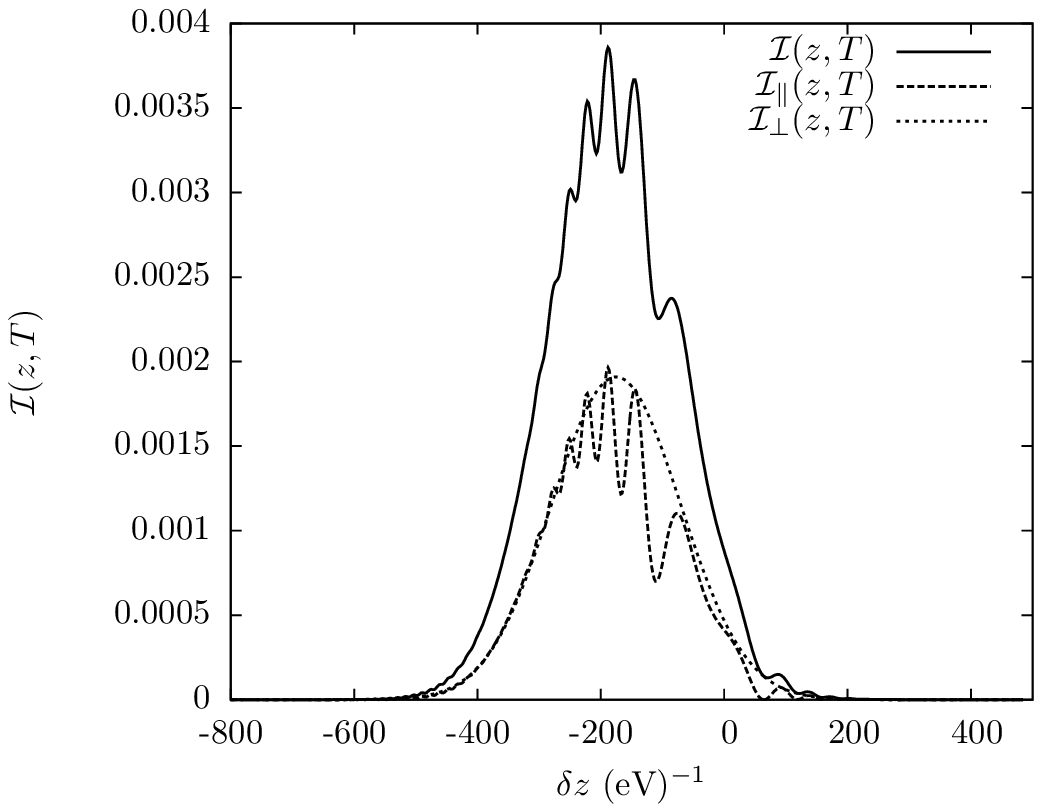}					\includegraphics[width=0.49\textwidth]{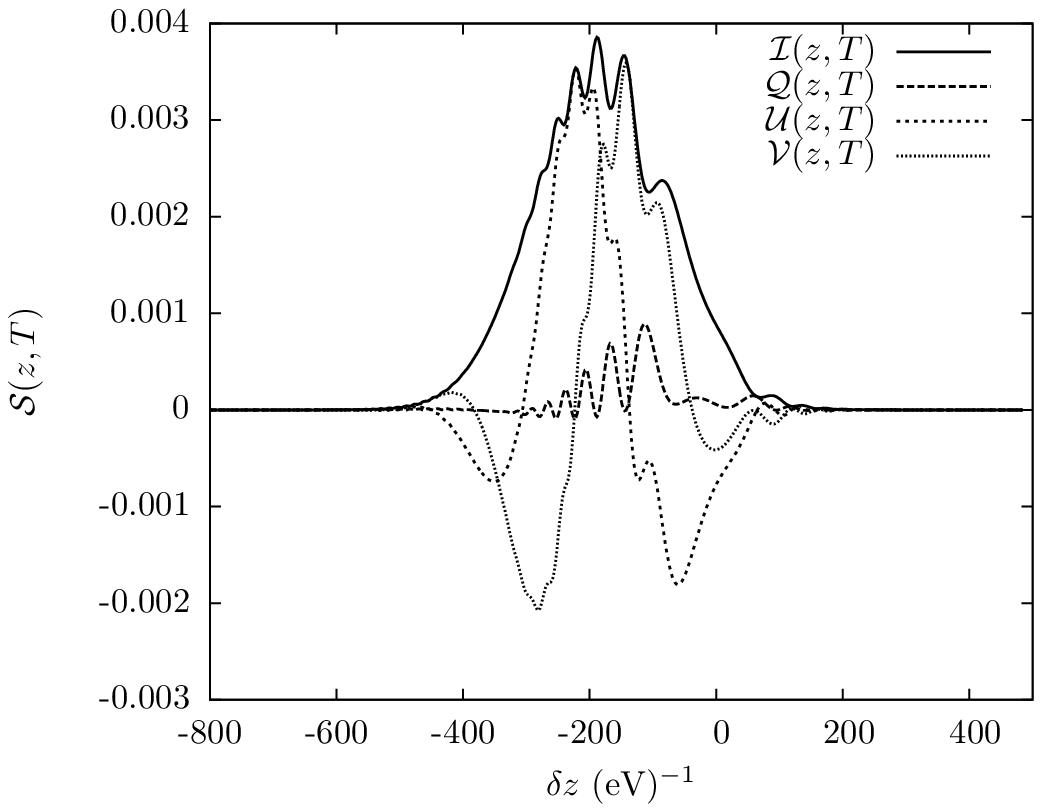}
			\end{center}
			\caption{The shape of the wave packets at time $T=10~\textrm{Mpc}/c$ for a light beam with $u(0)=1$. The abscissa is $\delta z\equiv z-cT$, which is the shift in position with respect to a frame moving at the speed of light $c$; \textit{i.e.} here the origin is at 10~Mpc. \emph{Left}: we show the total intensity and the intensities for the polarisations parallel and perpendicular to the magnetic field, before integration. \emph{Right}: we show the contributions to the other Stokes parameters. We used $\omega_p=3.7~10^{-14}$~eV, $m=4~10^{-14}$~eV, $\omega_0=2.5$~eV (\textit{i.e.} $\lambda_0=500$~nm), $a=1.34$~eV$^{-1}$, and $g\mathcal{B}=3~10^{-29}$~eV.
			}
			\label{fig:packets}
		\end{figure}

	\subsubsection{Stokes parameters and partially-polarised light}

	As wave packets go through the detector much faster than its time resolution, one has to integrate the packets over the exposure time $\Delta t$
 to calculate the Stokes parameters. Let us now represent by $S$ any of the four Stokes parameters; with the notations of Eq.~\eqref{eq:StokesE}, the observed quantities are then:
	\be
		{S}(z) = \langle \mathcal{S}(z,t) \rangle \equiv \frac{1}{N^2} {\int}_t^{t + \Delta t} dt\ \mathcal{S}(z,t),
	\ee
	where we introduce the normalisation\footnote{This value of $N$ corresponds to an initial intensity of 1~eV$^{4}$.} constant $N=(2\pi)^{-3/4}\sqrt{a}$, which cancels in polarisation degrees and normalised Stokes parameters.
	In Fig.~\ref{fig:packets}, we illustrate the packets after a propagation inside region II in a strong mixing case.\footnote{We use the Multiple-Precision Floating-point library with correct Rounding~\cite{mpfr}.} For photon polarisations parallel to $\vec{\mathcal{B}}$ we see the effect of interferences within the packet, while there is only a spread for non-mixing photons.  Note that this is for 100\% polarised light, so that the obtained $U(z)$ and $V(z)$ are much bigger than what the same conditions would give for typical quasars light.
	Now we need a correct description of what happens to initially unpolarised and partially linearly polarised light described with wave packets. 
\bigskip

		To treat partially polarised light, one can make use of a useful property of Stokes parameters in the case of fully polarised light, defined as in Eq.~\eqref{eq:fullypollight}. For fixed external conditions, calculating Stokes parameters in the $\varphi_0=\frac{\pi}{4}$ case gives us access to the following quantities:
		\bea
                \left.
                	\begin{array}{llll}
				Q_{pol}(z;\varphi_0=\frac{\pi}{4}) = \langle \frac{1}{2} \left({|E_{\perp}|}^2 - {|E_{\parallel}|}^2\right) \rangle &\equiv& C_1(z), \\
				\ \!\ \!I_{pol}(z;\varphi_0=\frac{\pi}{4}) = \langle \frac{1}{2} \left({|E_{\perp}|}^2 + {|E_{\parallel}|}^2\right) \rangle &\equiv& C_2(z), \\
				U_{pol}(z;\varphi_0=\frac{\pi}{4}) = \ \ \ \langle 2\ \operatorname{Re}\{E_{\parallel}E^*_{\perp}\} \rangle &\equiv& C_U(z), \\
				V_{pol}(z;\varphi_0=\frac{\pi}{4}) = \ \ \ \langle 2\ \operatorname{Im}\{E_{\parallel}E^*_{\perp}\} \rangle &\equiv& C_V(z), 
                	\end{array} \right. 
			\label{eq:lesC}
		\eea
		which evolve with $z$ due to the mixing with pseudoscalars.
		One can easily show that the evolution of the Stokes parameters for any other light beam $\vec{\mathcal{E}}_r$, with initial angle $\varphi_0$, in the same conditions is:
                \bea
                \left\{
                	\begin{array}{llll}
                	        \ \!I_{pol}(z;\varphi_0) &=& C_1(z) \cos(2\varphi_0) + C_2(z)\\
                	        Q_{pol}(z;\varphi_0) &=& C_2(z) \cos(2\varphi_0) + C_1(z)\\
                	        U_{pol}(z;\varphi_0) &=& C_U(z) \sin(2\varphi_0)\\
				V_{pol}(z;\varphi_0) &=& C_V(z) \sin(2\varphi_0).
                	\end{array} \right., 
			\label{eq:Stokesangle}
                \eea
		\textit{i.e.} it is sufficient to calculate the coefficients~\eqref{eq:lesC}.

		Now, as we have already discussed, unpolarised light can be thought of as the average over every possible initial angle $\varphi_0$. 
Applied to Eq.~\eqref{eq:Stokesangle}, this averaging gives, for unpolarised light: $I_{unpol}(z) = C_2(z)$, $Q_{unpol}(z) = C_1(z)$ and $U_{unpol}(z)=V_{unpol}(z)=0$.

		The evolution of any Stokes parameter $S$ for a light beam initially with a partial linear polarisation, characterised by a given value of $p_{lin,0}$ and a value of $\varphi_0$, then follows:
		\be
			S_{partial}(z;\varphi_0,p_{lin,0})
			=
			p_{lin,0}\ S_{pol}(z;\varphi_0) + \frac{1}{2\pi} {\int}_0^{2\pi} (1 - p_{lin,0}) S_{pol}(z;\varphi) \ d\varphi
			\label{eq:Stokesanglepartialexplicit}
		\ee
		this finally leads to a natural generalisation of the fully polarised case:
                \bea
                \left\{
                	\begin{array}{llll}
                	        \ \!I_{partial}(z;\varphi_0,p_{lin,0}) &=& p_{lin,0}[C_1(z) \cos(2\varphi_0)] + C_2(z)\\
                	        Q_{partial}(z;\varphi_0,p_{lin,0}) &=& p_{lin,0}[C_2(z) \cos(2\varphi_0)] + C_1(z)\\
                	        U_{partial}(z;\varphi_0,p_{lin,0}) &=& p_{lin,0}[C_U(z) \sin(2\varphi_0)]\\
				V_{partial}(z;\varphi_0,p_{lin,0}) &=& p_{lin,0}[C_V(z) \sin(2\varphi_0)].
                	\end{array} \right. 
			\label{eq:Stokesanglepartial}
                \eea

	\subsection{Results for white light}\label{section:results1zone}

		\begin{figure}
			\begin{center}
					\hspace{3cm}\includegraphics[width=0.7\textwidth]{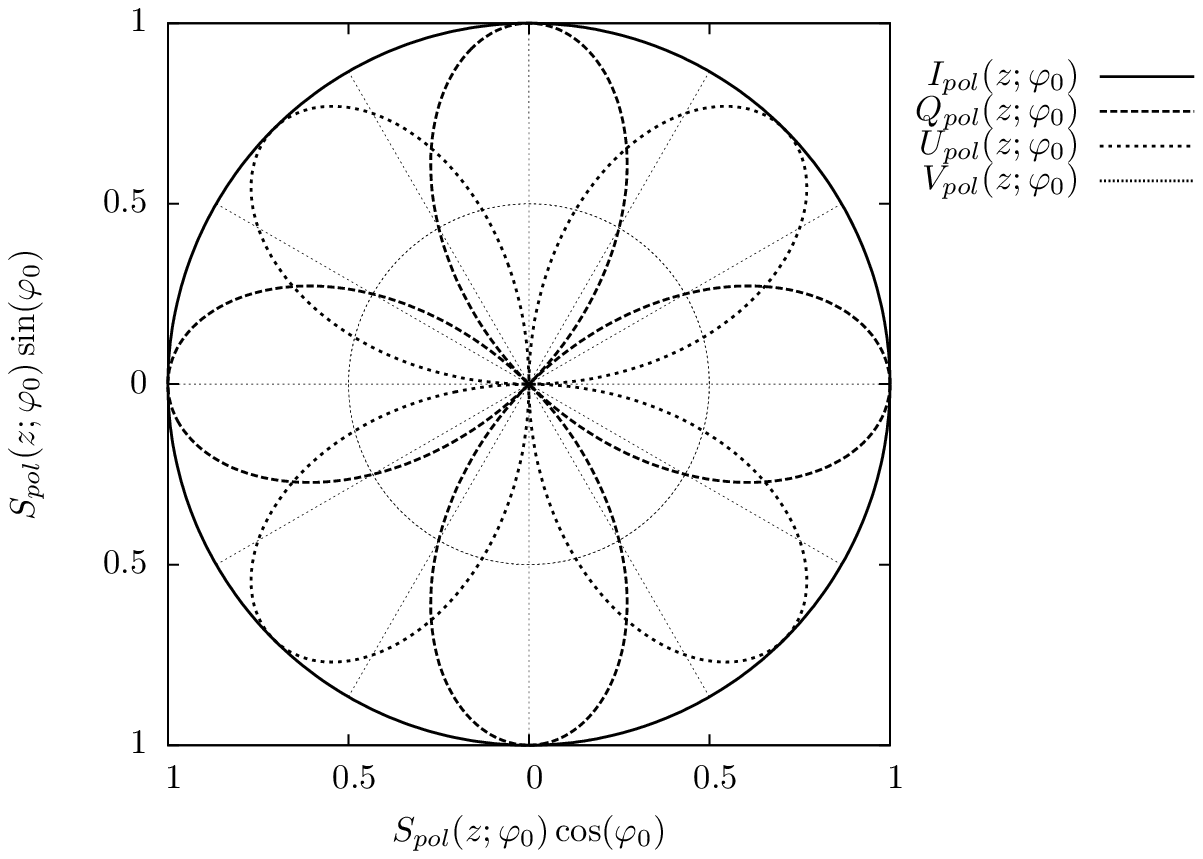}

					\includegraphics[width=0.55278\textwidth]{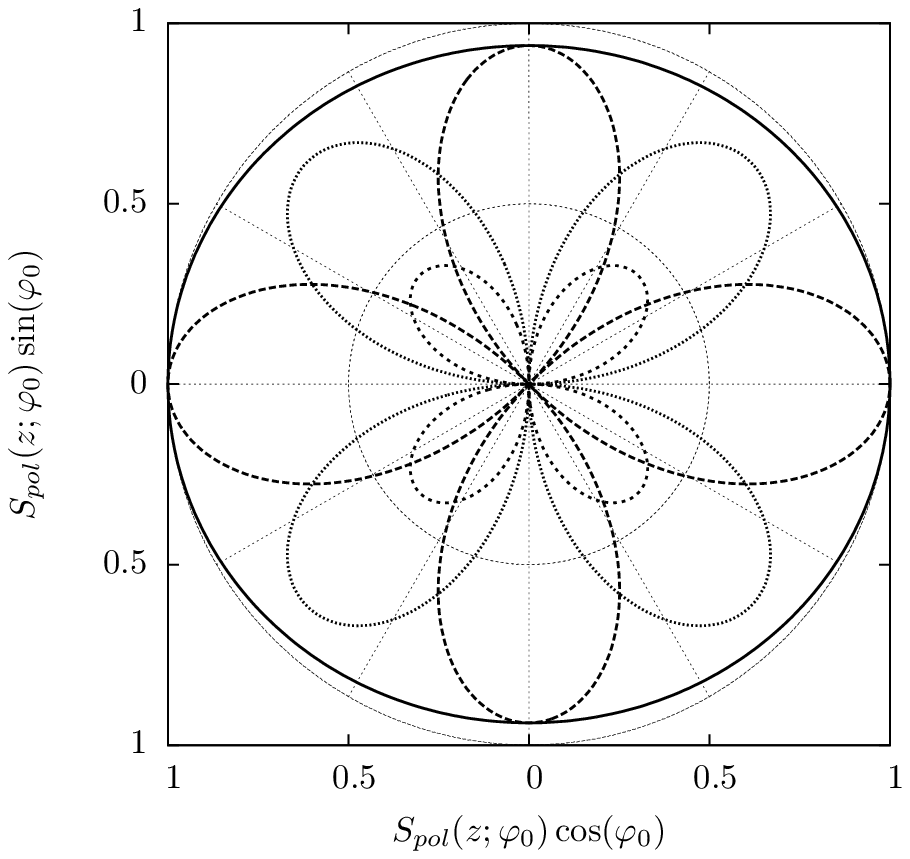}\includegraphics[width=0.45722\textwidth]{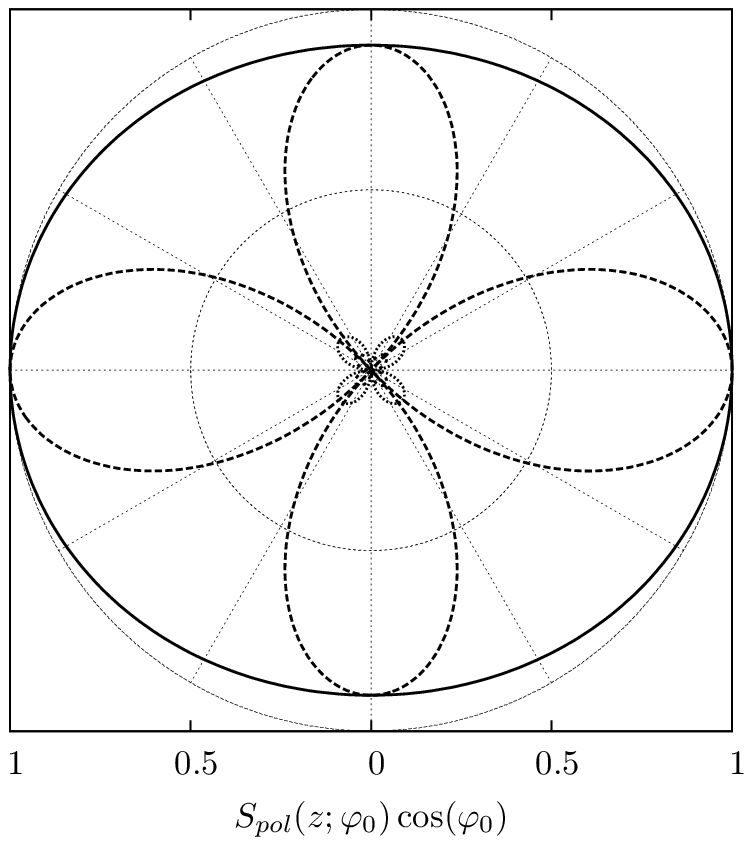}
			\end{center}
			\caption{Stereographic views of each of the Stokes parameters before (\emph{top}), and after a 10~Mpc propagation inside a magnetic field with plane waves (\emph{bottom left}), and with wave packets (\emph{bottom right}), for initially 100\% linearly polarised light. The distance of the curves to the origin gives the value of the parameters. To enable direct comparisons, the angular coordinate in the three figures is the initial angle, $\varphi_0$. The direction of the magnetic field is the one given by $\varphi_0=\pm\frac{\pi}{2}$. This relatively strong mixing case is shown for $m=4.5~10^{-14}$~eV, $\omega_p=3.7~10^{-14}$~eV, $g\mathcal{B}=5~10^{-29}$~eV, $\omega_0=2.5$~eV and $a = 1.34$~eV$^{-1}$.
			}
			\label{fig:polarstokes}
		\end{figure}

		\begin{figure}[h!!]
			\begin{center}
					\includegraphics[width=0.8\textwidth]{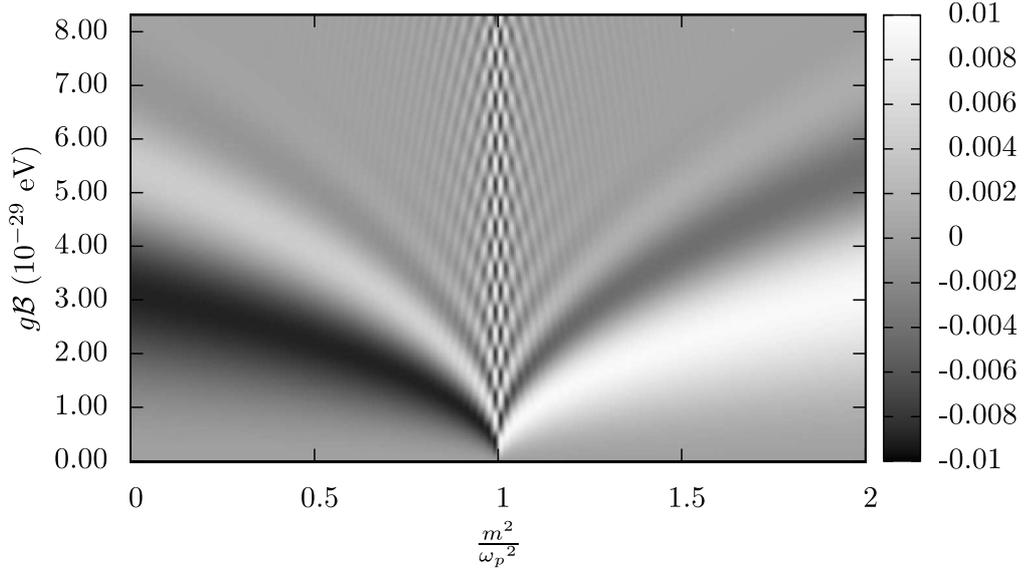}
			\end{center}
			\caption{Same (circular polarisation) as Fig.~\ref{fig:IPpcirc_pl} but with light described by wave packets. We have used $\omega_0=2.5$~eV and $a=1.34~$eV$^{-1}$.
			}
			\label{fig:WPpcirc}
		\end{figure}

		\begin{figure}[h!!]
			\begin{center}
					\includegraphics[width=0.8\textwidth]{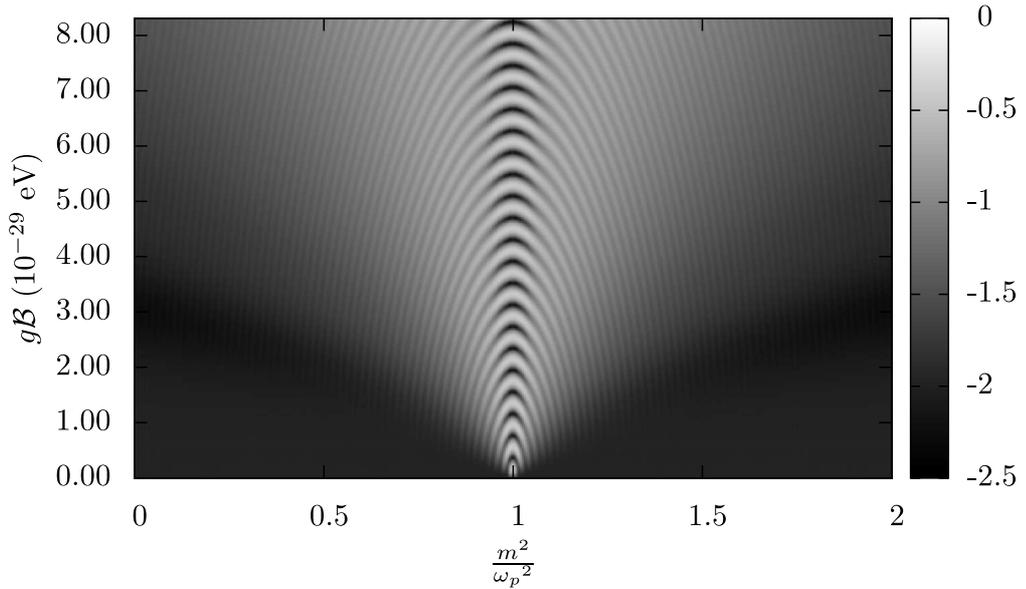}
			\end{center}
			\caption{Same (linear polarisation) as Fig.~\ref{fig:IPplin_pl} but with light described by wave packets. We have used $\omega_0=2.5$~eV and $a=1.34~$eV$^{-1}$. Note that the right-hand box gives the base-10 logarithm of the linear polarisation.
			}
			\label{fig:WPplin}
		\end{figure}

		\begin{figure}[h!!]
			\begin{center}
					\includegraphics[width=0.8\textwidth]{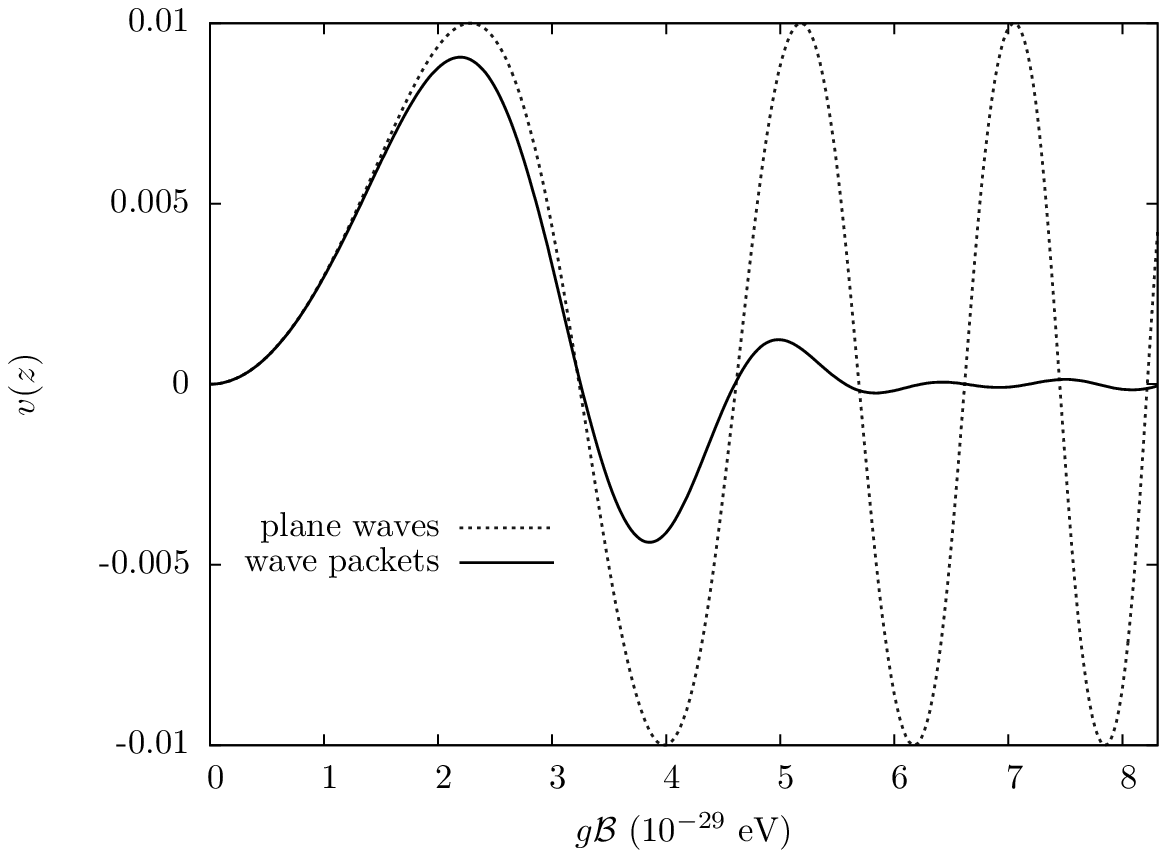}
			\end{center}
			\caption{Comparison of results obtained with plane waves and with wave packets, for the same parameters. This is a cut respectively of Fig.~\ref{fig:IPpcirc_pl} and of Fig.~\ref{fig:WPpcirc}, for the pseudoscalar mass $m=4.5~10^{-14}$~eV.
			}
			\label{fig:pcirc_compared}
		\end{figure}

	We now present the results of the mixing of photons with axion-like particles in a wave-packet formalism. We shall argue that these packets can be used to describe white light, with no photometric filter. The photomultipliers used to perform the white-light measurements of polarisation have a broad spectral response range (from 185 to 930~nm for the ones used in~\cite{Landstreet:1972so}), which is indeed similar to the width of our wave packets.

	Note that current upper limits on the pseudoscalar\footnote{Interestingly, for chameleons constraints are even less severe~\cite{Burrage:2009mj}.} coupling are $g=10^{-11}$~GeV$^{-1}$. Together with $\mathcal{B}=0.3~\mu$G, this means that $g\mathcal{B}\lesssim6~10^{-29}$~eV.

	In Fig.~\ref{fig:polarstokes}, we first illustrate the different Stokes parameters at a given distance, for each initial value of the angle $\varphi_0$. For a given angle, the distance between the origin and each $S(z;\varphi_0)$ curve is the value of this Stokes parameter. Now, for wave packets, we obtain that the circular polarisation $V(z)$ is strongly reduced with respect to the plane-wave prediction; notice that $U(z)$ is also affected in the same way, as it is also very sensitive to phase effects.
This can be understood if one goes back to Fig.~\ref{fig:packets}: we see that these two quantities change sign within the packet itself, due to the extremely frequency-dependent character of the birefringent effect induced by the pseudoscalars.
Also note that, whereas for plane waves the Stokes parameters obey $I^2(z)=Q^2(z)+U^2(z)+V^2(z)$ for any $\varphi_0$, this is no longer true in the wave-packet case, even for light initially fully linearly polarised.

	For partially-polarised light the expected amount of circular polarisation will of course be even smaller.
	This is shown in Figs.~\ref{fig:WPpcirc} and~\ref{fig:WPplin}, which are the wave-packet results analogous to those of the plane-wave case (Figs.~\ref{fig:IPpcirc_pl} and~\ref{fig:IPplin_pl}). We obtain a large suppression of circular polarisation for most of the parameters. Let us emphasise that this is in the case leading to the highest amount of $v$; \textit{i.e.} $u(0)=p_{lin}(0)=0.01$. Besides, notice that the maximum linear polarisation attainable for some of the parameters is smaller than in the plane-wave case: this is related to the loss of $u$ that happens in this case, as also seen in Fig.~\ref{fig:polarstokes}.

	In Fig.~\ref{fig:pcirc_compared}, we also directly compare the two descriptions for different values of the coupling; one can see that, for very small $g\mathcal{B}$, the results are similar, and that the suppression is more efficient at bigger values of $g\mathcal{B}$.
	\bigskip

	Finally, we have generalised our calculations to the case where the packets are initially described by the frequency distribution~\eqref{eq:gaussianpacket}, somewhere in the first region, at some $\tilde{z}\ll0$. The first region can then represent a cosmic void, where $\omega_p$ can also typically have a smaller value; this allows the packet to propagate a long time, which makes it spread, before it enters the second region. We have checked that the results we have presented above hold in this case as well (even if the first region is taken to be one gigaparsec long). This confirms that the main mechanism that reduces the circular polarisation is not related to the separation and the spread of photon packets of different polarisation, but rather because of phase-shifts within the packets that mix.
	This can be understood as $v(\omega)$ can change sign within the packet, averaging to zero, while $p_{lin}(\omega) = \sqrt{q^2(\omega) + u^2(\omega)}$ cannot, keeping an alignment possible.

\begin{figure}[h!!]
			\begin{center}
					\includegraphics[width=0.4825\textwidth]{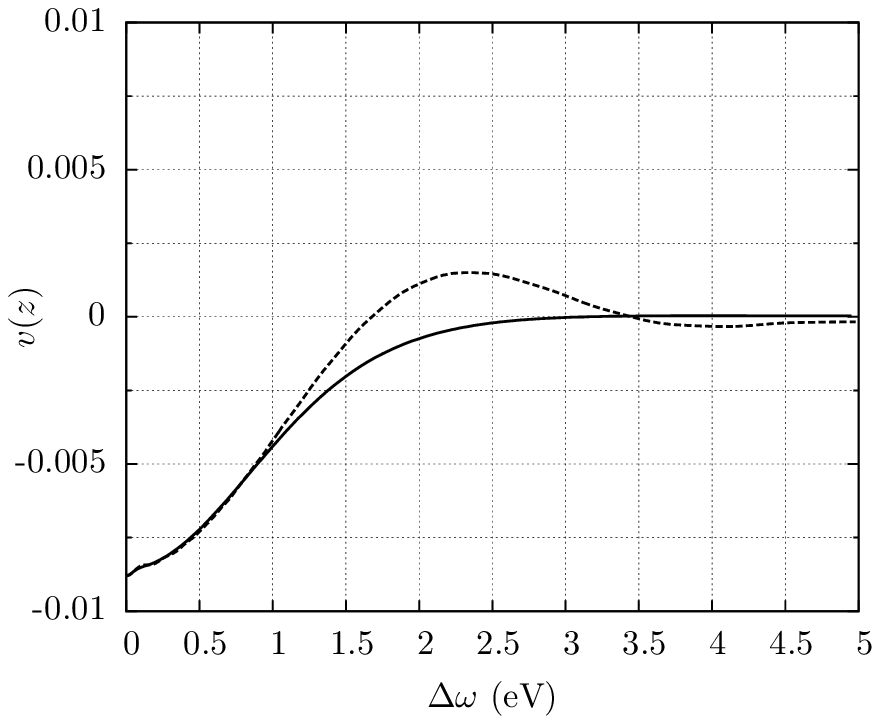}
					\includegraphics[width=0.508\textwidth]{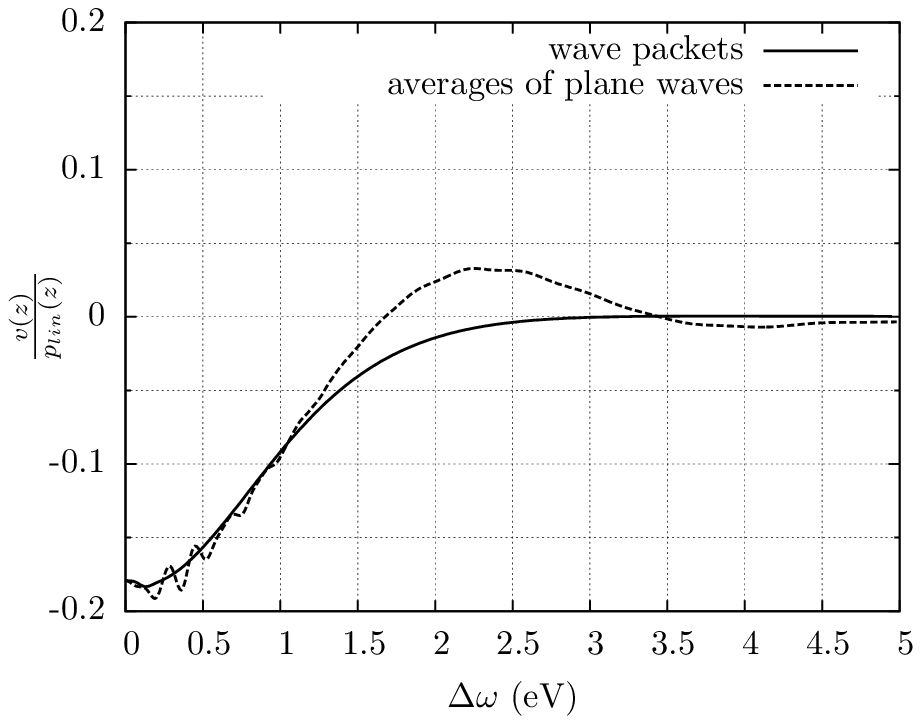}
			\end{center}
			\caption{Comparing averaging methods for wave packets (with $a=\frac{4\sqrt{\ln(2)}}{\Delta\omega}$) and averages of plane waves: in both cases, the absolute values of $v$ (\emph{left}) and of $\frac{v}{p_{lin}}$ (\emph{right}) are reduced with increasing $\Delta \omega$ with respect to the monochromatic case (\textit{i.e.} $\Delta \omega = 0$). Here, we used $g\mathcal{B} = 6~10^{-29}$~eV, and $\Delta \omega$ is centered around $\omega_0 = 2.5$~eV; the other parameters are the same as in Fig.~\ref{fig:pcirc_compared}.
			}
			\label{fig:av_vs_pckts}
		\end{figure}

	A simpler approach~\cite{Stodolsky:1998tc} is to use direct averages of the plane-wave Stokes parameters of Eq.~\eqref{eq:Stokes_alternative} over frequency, instead of wave packets. This will give the same qualitative results, as illustrated in Fig.~\ref{fig:av_vs_pckts}, where quantities are plotted against $\Delta \omega$, the bandwidth over which each averaging is performed. For the averages of plane waves, we have used the analytical formulas~\eqref{eq:Stokes_alternative} averaged over a step profile in $\omega$, centered around $\omega_0$ and of width $\Delta \omega$. For the Gaussian wave packets of Eq.~\eqref{eq:gaussianpacket}, on the other hand, we chose $\Delta \omega$ to represent the full-width-at-half-maximum in $\omega$ of each initial packet (\textit {i.e}, $a=4\sqrt{\ln(2)}{\left(\Delta\omega\right)}^{-1}$). In either case, the larger the band of frequencies over which the averaging is done, the smaller the absolute value of the circular polarisation and its relative importance compared to the linear polarisation. This holds whatever the details of the averaging. Similar results have also been obtained in different contexts (chameleons~\cite{Burrage:2008ii}, and high-energy gamma sources~\cite{Bassan:2010ya}).

	Therefore, as far as white-light data are concerned, phenomenological implications of axion-like particles mixing with photons can be reconciled with circular polarisation measurements~\cite{Payez:2009kc,Payez:2009vi,Payez:2010xb}.
	
	\subsection{Results for Bessell V-filter}\label{sec:vfilterKO}

	Most of the recent circular polarisation data of~\cite{Hutsemekers:2010fw} were taken using a Bessell broadband V filter~\cite{Sterken:1992ap,EFOSC2:2008}. This filter is centered around $\lambda = 547.6$~nm and the associated full-width-at-half-maximum is 113.2~nm. To mimic this cut in frequencies, one can convolute wave-packets with the spectrum distribution of the filter, or proceed to averages of plane-wave results over $\omega$ using the frequency profile.

	We then find that, even though it is a broadband filter, the typical values of the astrophysical parameters are such that the circular polarisation does not change sufficiently over this bandwidth to be strongly reduced when averaging over $\omega$. This is illustrated in Fig.~\ref{fig:av_vs_pckts} for small values of $\Delta \omega$ ($\approx 0.5$~eV). The circular polarisation is slightly smaller than in the monochromatic case, but the effect is certainly not sufficient to reconcile the mechanism with data. Except for very specific choices of parameters, the axion-like particle parameters able to create an alignment will also predict a sizeable amount of circular polarisation.

	If axions were at work, given the ---somehow narrow--- bandwidth of the broadband V-filter, circular polarisation should have been observed, even with a wave-packet description.

	\section{Different models for the magnetic field}\label{sec:ruledout}

	We are going to focus on the regions where these V-filter data have been taken from and check the sensitivity of this result to changes in the magnetic field morphology.
	We will see that the mixing with pseudoscalars in a magnetic field that fluctuates along the light trajectory cannot even reproduce the alignments of linear polarisation.

	\subsection{Linear polarisation data towards region A1}

	In order to detect alignments, we shall first argue that averages in the $(q,u)$ space give the same information as the more elaborate methods of~\cite{Hutsemekers:2005}.

	Among quasars with circular polarisation measured in the V-filter, 18 are located in the same direction of the sky, towards what is called region A1 in~\cite{Hutsemekers:1998,Hutsemekers:2001,Hutsemekers:2005}.
	We can present in a $(q,u)$ space the linear polarisation of quasars located in this direction\footnote{Doing so, we drop the information about the position of each object on the sky.}, and see what the alignment effect looks like in such a plot. 
	As the polarisation angle is related to the Stokes parameters $q$ and $u$ via the relation:
	\be
		\varphi= \frac{1}{2}\textrm{atan}\left(\frac{u}{q}\right),
	\ee
	for a fixed value of $p_{lin}$, different values of $q$ and $u$ correspond to different orientations. In particular, a random distribution of polarisation angles corresponds to an isotropic distribution in this space.
	Note also that, as $p_{lin} = \sqrt{q^2+u^2}$, the distance between the origin and a given point directly gives the degree of polarisation of the associated light beam.

	\begin{figure}[h!!]
		\begin{center}
			\includegraphics[width=0.5305\textwidth]{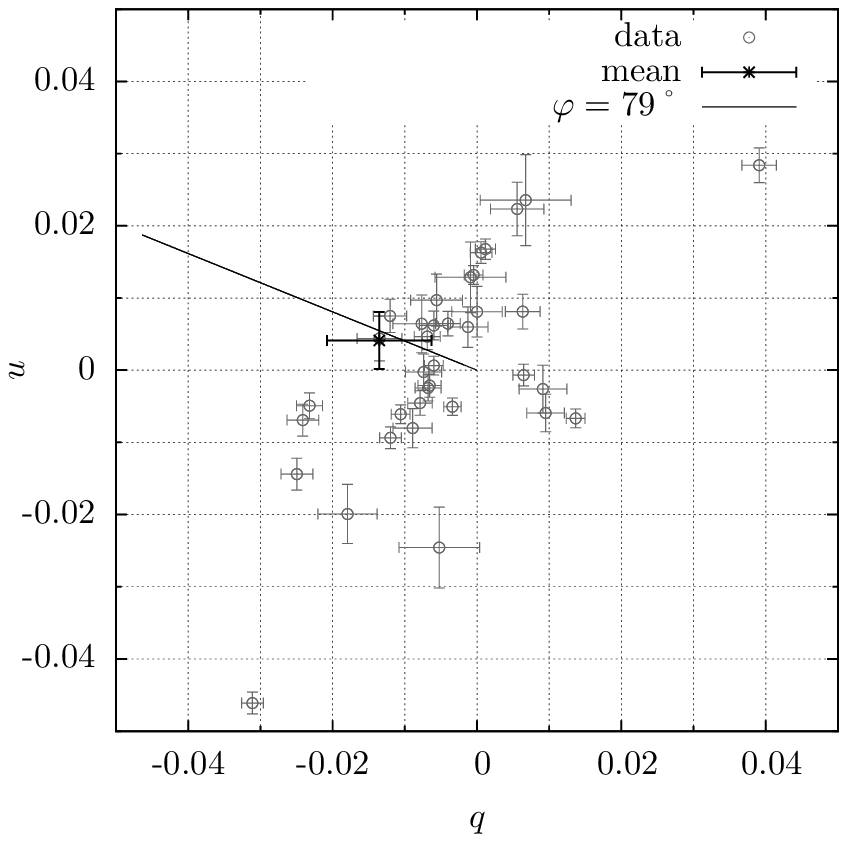}
			\includegraphics[width=0.46\textwidth]{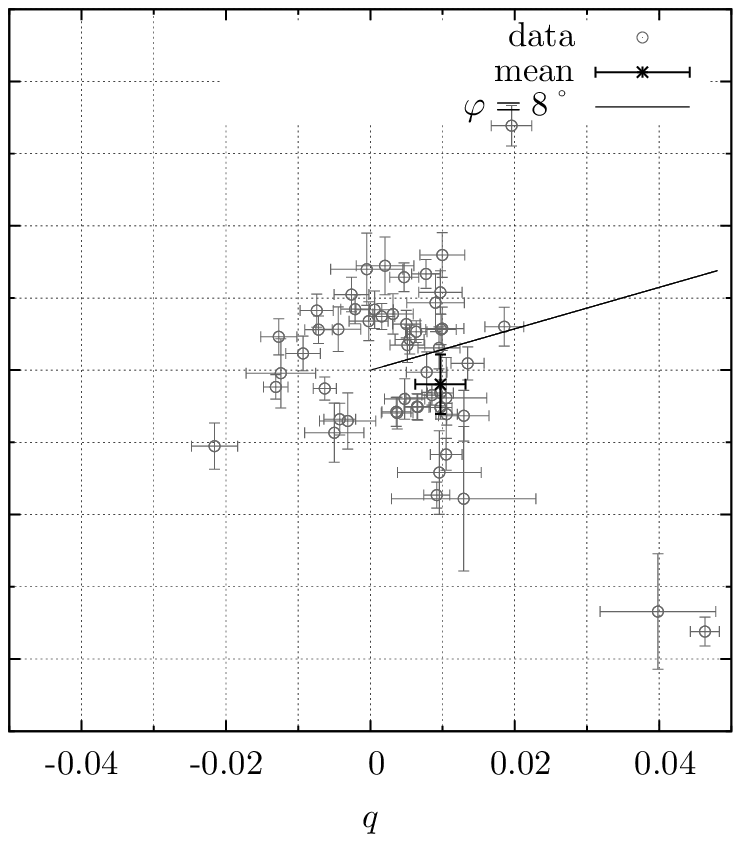}
		\end{center}
			\caption{Experimental data: objects taken in the A1 direction. \emph{Left}: linear polarisation for low-redshift quasars, with $0\leq z<1$. \emph{Right}: same for high-redshift ones, with $1\leq z \leq 2.3$. Some objects with higher polarisation degrees are not shown, but are taken into account for the mean. In the case of low-redshift quasars, the total number of objects is $N=43$, and for high-redshift ones, it is $N=56$.
			}
			\label{fig:q_u_data}
	\end{figure}

In Fig.~\ref{fig:q_u_data}, we show data from the latest sample~\cite{Hutsemekers:2005}, for both low- and high-redshift quasars.\footnote{Only objects with $p_{lin}\geq0.6\%$ have been considered in this sample, this is why there is a hole in the center of those figures.} 
As expected, coherent orientations are translated into departures from isotropy on such a graph. 
	Note that the preferred direction for the asymmetry is not the same for low and high redshifts, while these objects are along the same line of sight. 
\bigskip

	To be more quantitative, we can calculate the mean values of $q$ and of $u$, for low- and high-redshift data, taking into account the experimental uncertainties\footnote{For this, one takes $\sigma_q = \sigma_u = \sigma_p$, see Ref.~\cite{Serkowski:1958sa}}.
	We determine the mean values and the errors on the mean for $q$ and $u$ and plot them in Fig.~\ref{fig:q_u_data}; we obtain $(-0.0135\pm0.0072, 0.0041\pm0.0039)$ for the low-redshift region, and $(0.0097\pm0.0035, -0.0019\pm0.0041)$ for the high-redshift one. In the observational paper~\cite{Hutsemekers:2005}, another analysis was done: they obtained the preferred angles one finds when considering only the angular information; these are shown with straight lines.

	\subsection{Models}

	The location of region A1 points towards the center of the Virgo supercluster (our local supercluster, shortened as `LSC'). To the best of our knowledge of the Virgo magnetic field can be described either~\cite{Vallee:2002,Vallee:2011}:
	\begin{itemize}
		\item[-] by a uniform field (used thus far for illustration) with $\approx$0.2--0.3~$\mu$G over  $\approx$5--10~Mpc;
		\item[-] by a `patchy' field made by several $\approx$100~kpc cells of randomly oriented magnetic field of strength $\approx$2~$\mu$G, adding up to the same distance.
	\end{itemize}
Note that a `patchy' picture is also typically what is considered to discuss the propagation of cosmic rays, and what is obtained from structure formation; see for instance~\cite{Dolag:2003ft,Dolag:2005a} for results about the LSC.

	The LSC magnetic field is essentially the last relevant magnetic field encountered by extragalactic photons coming towards us.\footnote{The influence of our galactic magnetic field can be neglected: the field strength decreases exponentially in the direction transverse to the galactic plane~\cite{Giovannini:2003yn}, and data have been obtained at high galactic latitudes.} For this reason, regardless of its structure, the axion-like particle explanation of quasar data will be ruled out if the influence of this field creates too much circular polarisation, as any $v$ created there should have been detected.

	\subsubsection{Simulations in the uniform field scenario}

	We know since Section~\ref{sec:vfilterKO} that a uniform field can produce coherent orientations of polarisation with respect to the magnetic field direction, although the existence of axion-like particles responsible for an alignment would have also implied that circular polarisation is produced. As we have discussed, this is excluded by data.

	We can check what this alignment looks like in a $(q, u)$ space, if we start from a random distribution of polarisation. 

	\begin{figure}[h!!]
		\begin{center}
			\includegraphics[width=0.49\textwidth]{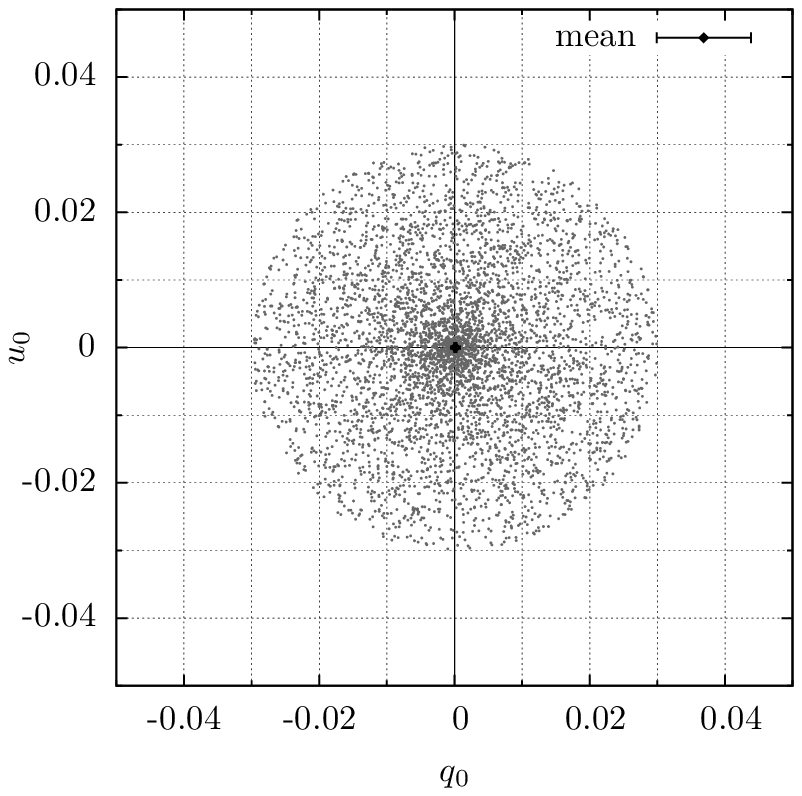}
			\includegraphics[width=0.499\textwidth]{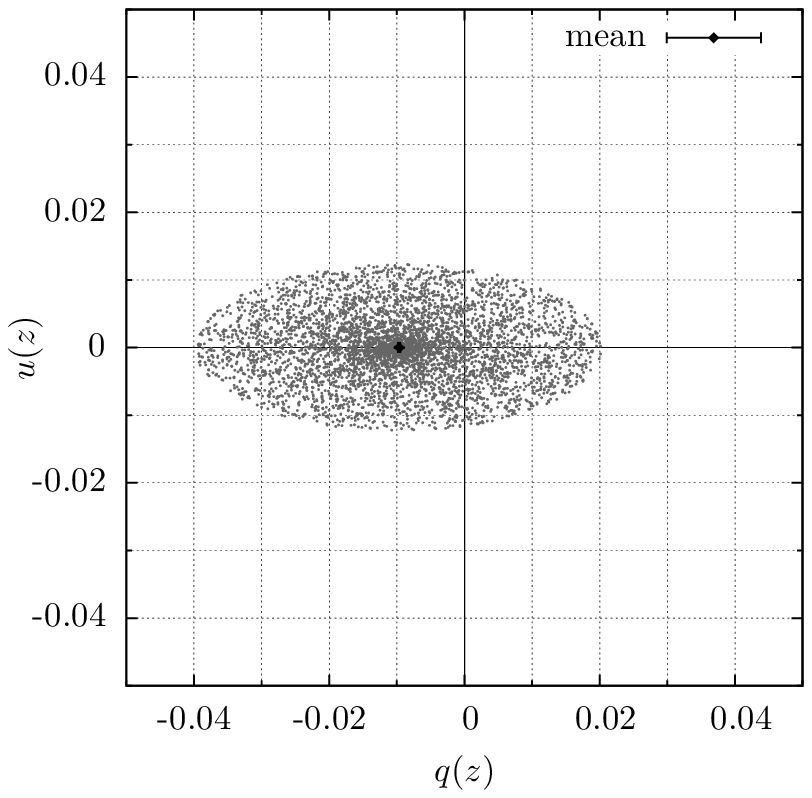}
		
		\end{center}
			\caption{5000 beams are generated. \emph{Left}: initial distribution. \emph{Right}: The associated distribution after effects induced by axion-photon mixing in the uniform case. The parameters used here are $\omega_0=2.25$~eV, $\omega_p=3.7~10^{-14}$~eV, $m=4.5~10^{-14}$~eV, $g = 3.5~10^{-12}$~GeV$^{-1}$, $\mathcal{B}=0.3~\mu$G, and $z=10$~Mpc.
			}
			\label{fig:q_u_uniform}
	\end{figure}

	To mimic a random distribution of initial quasar polarisations, we first generate partially polarised light beams ($p_{lin}$ between 0 and 3\%), with random polarisation angles. In Fig.~\ref{fig:q_u_uniform}, on the left, we plot this initial distribution of light beams, each random realisation being displayed using its Stokes parameters $q$ and $u$.
	On the right, we show what this distribution becomes, due to axion-photon mixing inside the 10~Mpc uniform magnetic field. We see that there is indeed a departure from a random distribution acquired through the mixing, corresponding to an asymmetry in the $q$ and $u$ space.\footnote{The fact that the asymmetry appears along one of the axes is only due to our specific choice for the basis; only $p_{lin}$ is a physical quantity.} 
More quantitatively, the means we obtain for $q$ and $u$ in this example lead to a value of $p_{lin}=0.01$ after axion-photon mixing, while they were compatible with zero initially.

	\subsubsection{Simulations in the patchy scenario}
	\paragraph{Pure randomness}
	As already mentioned, making the magnetic field fluctuate in a `patchy' model may suppress $v$: as we have seen in Section~\ref{sec:newcircpoldata}, in small enough magnetic field regions, the induced circular polarisation can be smaller than the linear one.
Nevertheless, circular polarisation is not the main problem in this picture.

Indeed, it is obvious that such a field will not help create an alignment: if the magnetic field can be thought of as small cells with magnetic field directions distributed in a random way from cell to cell, this will be the case along the line of sight, but also transversally. Then, two objects which are angularly separated will pass through two different magnetic field configurations. There is thus no way to create an alignment, as there is no preferred direction in this problem that is common to all quasars.

	\paragraph{With an underlying uniform field}

	We can go further and use a more refined model, where there would be at least some correlation between the cells rather than a complete randomness. To do this, we sum the magnetic fields of the uniform and of the `patchy' models, using results presented in Appendix~\ref{app:cells}. We thus have cell-like magnetic fields on top of a fainter field, this one being coherent over the LSC scale: this would lead to some semi-randomness between the cells on the LSC scale, as discussed in~\cite{Vallee:2002}.
	\begin{figure}[h]
		\begin{center}
			\includegraphics[width=0.5\textwidth]{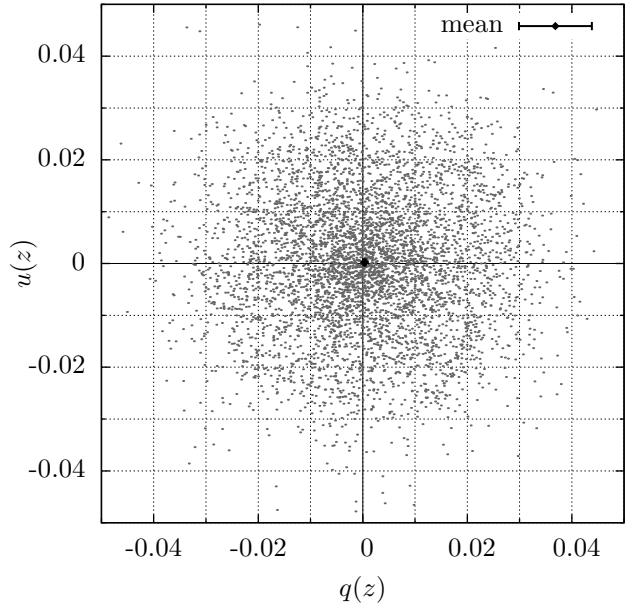}
		\end{center}
			\caption{
			Same as Fig.~\ref{fig:q_u_uniform} (\emph{right}) but after axion-photon mixing in the `patchy' case with an underlying uniform field. The parameters used here are the same as before, except $m=1.85~10^{-13}$~eV. For the uniform magnetic field, we use $\mathcal{B}=0.3~\mu$G over 10~Mpc; for the randomly oriented part of the magnetic field, we use $|\vec{\mathcal{B}}_c| = 2~\mu$G in a hundred $100~$kpc-long cells.
			}
			\label{fig:q_u_patchyplusundl}
		\end{figure}

	In this case, to keep a final linear polarisation of the order of 1\%, we have to consider either smaller values of the coupling, or bigger values of $|(m^2 - {\omega_p}^2)|$ than in the uniform case because the magnetic field is stronger (see Eq.~\eqref{eq:thetamix}).

	For this reason, and because the field strength of the uniform component is $\approx7$ times smaller than that of the randomly oriented one, it is not surprising that there is no obvious departure from isotropy due to axion-photon mixing in this case. Indeed, the effect induced by the uniform component of the magnetic field is then strongly suppressed: therefore, an alignment cannot be achieved.
	For the example we present in Fig.~\ref{fig:q_u_patchyplusundl}, we obtain that the means of $q$ and $u$ are compatible with zero at the 1$\sigma$-level, namely hardly any improvement with respect to the initial distribution, and certainly not enough additional linear polarisation to be able to explain the data. 
	We checked that if the relative intensities of the random and background fields are chosen to produce an alignment, then the circular polarisation is again too high.

	If this second possibility turns out to be a satisfactory model of the magnetic field of the local supercluster, not only would there be some circular polarisation, but axion-like particles will be unable to create coherent orientations in that field. Note that these results are general and do not only apply to the LSC magnetic field.\footnote{Note also that considering another field should imply that it is coherent over angular distances even larger than the LSC scale: indeed, it should be located beyond it and still large enough so that light from angularly distant quasars pass through a field giving the same preferred direction.}

	Finally, we have also checked that our results are stable with respect to fluctuations of a factor of 2 (up or down) of the parameters, among which the plasma frequency, the cell sizes, and the magnetic field strength.

	\section{Conclusion}

	In this paper, we have considered photon-pseudoscalar mixing as the source of the observed alignments. This process has so far been the best hope to explain the observed data. The typical treatment assumes that the same faint coherent field is traversed by the light beams from all the quasars. Note that this is problematic, as the data display two alignments: quasars at redshift $z>1$ seem to be aligned in a different direction from closer ones. In order to obtain such an effect, one needs to assume a slice of coherent fields 1~Gpc large, with an intensity of the order of 0.1~$\mu$G. Although this seems unlikely, one cannot rule out the explanation on those grounds, as so little is known about magnetic fields at high redshifts.

	Recent data have shown that the circular polarisation of these objects is negligible, contrarily to the predictions from photon-pseudoscalar mixing. At face value, this kills the interpretation. However, we showed that it is possible to argue that, for white light, the average over frequencies suppresses the circular polarisation enough. From the sample of 355 quasars, only 6 circular polarisation measurements were obtained in white light, and 21 using a broadband Bessell V-filter, for which the bandwidth is nevertheless not broad enough to average the circular polarisation to almost zero.

	Keeping the idea of averaging, we extended the model of magnetic field, by assuming a collection of cells of 100 kpc, with random 2~$\mu$G fields averaging to 0.3~$\mu$G, and letting most quantities fluctuate (size of cells, electron density, magnetic field). While these fluctuations can somewhat reduce the circular polarisation, they also destroy the alignment which was the first reason to consider photon-pseudoscalar mixing.

	Hence it seems that the combination of the alignments and of the absence of circular polarisation remains at present a puzzle, which cannot be explained by photon-pseudoscalar mixing. One should note that this conclusion is based on the lack of circular polarisation in 19 objects measured with the V-filter. It is of course possible that magnetic field configurations are special for these 19 objects, and some more data~---or, equivalently, data in an even smaller bandwidth---~may be needed to reach certainty.

	As a final note, the data on quasar polarisation show that the polarisations are only typically a few percent. Photon-pseudoscalar mixing can be much more efficient than this at producing polarisation. Hence the quasar data can be used to exclude part of the parameter space of ALPs. This will be the subject of a future paper~\cite{tocome}.

	\section*{Acknowledgements}

	A.~P. would like to thank Davide Mancusi for numerous discussions on technical issues, Fredrik Sandin for his help about the use of the MPFR library. A.~P. also thank Javier Redondo for interesting exchanges about axion physics, as well as C{\'e}dric Lorc{\'e}, Diego Aristizabal, and Federico Ceccopieri for stimulating discussions. Besides, A.~P. thank Georg Raffelt who suggested what became the end of Section~\ref{section:results1zone}.
	D.~H. is Senior Research Associate at F.R.S-FNRS; A.~P. is an IISN researcher.

\appendix

\section{Pseudoscalar-photon mixing using wave packets}\label{app:wp}

	Here, we solve the system of Eq.~\eqref{eq:system} in the case of the step-like magnetic field presented in Section~\ref{sec:wptreatment}, using the decomposition of Eq.~\eqref{eq:decompsuromega}.
	Note that, as in the plane-wave case, we rephase $\phi(z,t)$ and use the gauge condition $A^0=0$, so that $\widetilde{E}(z,\omega)=i\omega\widetilde{A}(z,\omega)$.

	The solutions in the first region are:
	\bea
		E_I (z,t) &=& \int_{-\infty}^{\infty} d\omega\   e^{- i \omega t} i\omega\big[\widetilde{A}_{i, I}(z = 0,\omega) e^{i k_E z} + \widetilde{A}_{r, I}(z = 0,\omega) e^{- i k_E z}\big]\\
		\phi_I (z,t) &=& \int_{-\infty}^{\infty} d\omega\   e^{- i \omega t} \big[\widetilde{\phi}_{i, I}(z = 0,\omega) e^{i k_{\phi} z} \ + \widetilde{\phi}_{r, I}(z = 0,\omega) e^{- i k_{\phi} z}\big],
	\eea
with the dispersion relations $k_E = \sqrt{\omega^2 - {\omega_p}^2}$ and $k_{\phi} =\sqrt{\omega^2 - m^2}$. Here we have already used the fact that we will always consider amplitudes centered around $\omega_0$, with $\omega_0\gg\omega_p, m$, and decreasing sufficiently quickly with $\omega$ for the contributions from $\omega\leq\omega_p, m$ to be negligible. Similarly, the solutions in the second region read:
        \begin{multline}
		E_{II} (z,t) = \int_{-\infty}^{\infty} d\omega\   e^{- i \omega t}   i \omega  \big[ \widetilde{A}_{i, II}(z = 0,\omega) \big((\textrm{c}_{mix})^2 e^{i k_C z} + (\textrm{s}_{mix})^2 e^{i k_D z}\big)\\
+  \widetilde{\phi}_{i, II}(z = 0,\omega) \frac{\sin(2\theta_{mix})}{2}\big(e^{i k_C z} - e^{i k_D z}\big)\big],
	        \label{eq:EIIxtevolution}
	\end{multline}
        \begin{multline}
		\phi_{II} (z,t) = \int_{-\infty}^{\infty} d\omega\    e^{- i \omega t} \big[ \widetilde{A}_{i, II}(z = 0,\omega)  \frac{\sin(2\theta_{mix})}{2}\big(e^{i k_C z} - e^{i k_D z}\big)\\
+  \widetilde{\phi}_{i, II}(z = 0,\omega) \big((\textrm{s}_{mix})^2 e^{i k_C z}  +  (\textrm{c}_{mix})^2 e^{i k_D z}\big)\big],
	        \label{eq:phiIIxtevolution}
	\end{multline}
	where $k_C$ and $k_D$ are respectively $k_+=\sqrt{\omega^2-{\mu_+}^2}$ and $k_-=\sqrt{\omega^2-{\mu_-}^2}$ when $\omega_p>m$, and the other way around when $m>\omega_p$.
	As we are mostly interested in the small mixing case, the heaviest eigenmode of propagation is mostly made of the heaviest state among photons and pseudoscalars, and conversely for the lightest one. If $\mathcal{B}$ is set to zero, $k_C=k_E$ and $k_D=k_{\phi}$.
\bigskip

	The amplitudes $\widetilde{A}_{i, I}(0,\omega)$, $\widetilde{A}_{r, I}(0,\omega)$, $\widetilde{\phi}_{i, I}(0,\omega)$, $\widetilde{\phi}_{r, I}(0,\omega)$, $\widetilde{A}_{i, II}(0,\omega)$ and $\widetilde{\phi}_{i, II}(0,\omega)$ are determined by initial and boundary conditions. They correspond to incident ($i$) or reflected ($r$) amplitudes that appear as light goes from region I into the potential barrier.
To simplify our discussion, we now work in the case where there is no incident pseudoscalar flux in region I, namely ${\phi}_{i, I}(0,\omega) = 0$.

	The continuity requirements on $E(z,t)$ and $\phi(z,t)$, and on their first derivatives with respect to $z$, at $z=0$ then lead to relations where the only free parameter left is $\widetilde{A}_{i, I}(z=0,\omega)$. For completeness, they are:
        \begin{multline}
		\widetilde{A}_{i, II} (z=0, \omega) = \\  2k_E \Big[   \frac{k_C (\textrm{s}_{mix})^2 + k_D (\textrm{c}_{mix})^2 + k_{\phi}}{k_E k_{\phi} + k_C k_D + k_E \big( k_C(\textrm{s}_{mix})^2 + k_D (\textrm{c}_{mix})^2 \big) + k_{\phi} \big( k_C (\textrm{c}_{mix})^2 + k_D (\textrm{s}_{mix})^2 \big)}   \Big] \widetilde{A}_{i, I}(0,\omega)\\
		\equiv \mathscr{V}\ \widetilde{A}_{i, I}(0,\omega)
        	\label{eq:AiIIcoeff}
	\end{multline}
        \begin{multline}
		\widetilde{\phi}_{i, II}(z=0,\omega) = \widetilde{\phi}_{r, I}(z=0,\omega) = \\
 \Big[   \frac{k_E \big(k_C - k_D\big) \sin(2\theta_{mix})}{k_E k_{\phi} + k_C k_D + k_E \big( k_C(\textrm{s}_{mix})^2 + k_D (\textrm{c}_{mix})^2 \big) + k_{\phi} \big( k_C (\textrm{c}_{mix})^2 + k_D (\textrm{s}_{mix})^2 \big)}   \Big] \widetilde{A}_{i, I}(0,\omega)\\
	\equiv \mathscr{W}\ \widetilde{A}_{i, I}(0,\omega).
	        \label{eq:phiiIIcoeff}
	\end{multline}
        \begin{multline}
		\widetilde{A}_{r, I}(z = 0,\omega) = \left(\mathscr{V} - 1\right) \widetilde{A}_{i,I}(0,\omega)\\
        \end{multline}

	In the case of an incident Gaussian wave packet
	\be
		\widetilde{{E}}_{i, I}(z = 0,\omega) = e^{-\frac{a^2}{4}(\omega-\omega0)^2},
	\ee
	we obtain the following result for $E_{II}(z,t)$ (which is the only amplitude entering the expressions of the Stokes parameters in the second region)%
:
	\begin{multline}
		 E_{II}(z,t) = \int_{-\infty}^{\infty} d\omega\ 
		 \Big[\frac{\mathscr{V} - i \mathscr{W}}{4} \exp(-\frac{a^2}{4}(\omega-\omega0)^2 + i(k_Cz - \omega t + 2\theta_{mix})) \\
		 + \frac{\mathscr{V} +  i \mathscr{W}}{4} \exp(-\frac{a^2}{4}(\omega-\omega0)^2 + i(k_Cz - \omega t - 2\theta_{mix})) \\
		 + \frac{\mathscr{V}}{2} \exp(-\frac{a^2}{4}(\omega-\omega0)^2 + i(k_Cz - \omega t)) \\
		 - \frac{\mathscr{V} - i \mathscr{W}}{4} \exp(-\frac{a^2}{4}(\omega-\omega0)^2 + i(k_Dz - \omega t + 2\theta_{mix})) \\
		 - \frac{\mathscr{V} + i \mathscr{W}}{4} \exp(-\frac{a^2}{4}(\omega-\omega0)^2 + i(k_Dz - \omega t - 2\theta_{mix})) \\
		 + \frac{\mathscr{V}}{2} \exp(-\frac{a^2}{4}(\omega-\omega0)^2 + i(k_Dz - \omega t))];\label{Eparallel_generalexplicit}
	\end{multline}
	where $\mathscr{V}$, $\mathscr{W}$, $k_C$, $k_D$ and $\theta_{mix}$ are functions of $\omega$. 
	Note that if $g\mathcal{B}$ is set to zero, this reduces to
	\be
		E_{II}(z,t;g\mathcal{B}=0)=E_{\perp}(z,t) = \int_{-\infty}^{\infty} d\omega\ \exp(-\frac{a^2}{4}(\omega-\omega0)^2 + i(k_Ez - \omega t)).
		\label{E_gbzero}
	\ee
We finally Taylor expand the coefficients and the arguments of the exponentials around $\omega_0$ (up to the second order) to carry out the integrals~\eqref{Eparallel_generalexplicit} and \eqref{E_gbzero} analytically to better than 1\% for the case at hand (as was checked by estimating the contribution of the next order).

\section{Mixing in a more general magnetic field}\label{app:cells}

	Consider several regions with different magnetic fields, their direction and strength changing from one region to another. First of all, one can work out axion-photon mixing in a arbitrarily-oriented transverse magnetic field $\vec{\mathcal{B}} = \mathcal{B}\cos\delta~\vec{e}_1 + \mathcal{B}\sin\delta~\vec{e}_2$, where $\vec{e}_1$ and $\vec{e}_2$ are the basis vectors we will use throughout to keep track of an absolute direction and to define Stokes parameters.

	We approximate $(\omega^2 + {\partial_z}^2) \simeq 2\omega\left(\omega + i\partial_z\right)$ in the equations of motion for the fields, as the masses we use are indeed much smaller than the photon energies entering the problem. Inside a region, the system of equations reads:
	    \be
		    \Bigg[\Big(\omega + i\frac{\partial}{\partial z}\Big) -
	                                \left(
	                                \begin{array}{ccc}
	                                 \frac{{\omega_p}^2}{2\omega} & 0            & \frac{-g\mathcal{B}\cos\delta}{2}\\
	                                0            &  \frac{{\omega_p}^2}{2\omega} & \frac{-g\mathcal{B}\sin\delta}{2}\\
	                                \frac{-g\mathcal{B}\cos\delta}{2}            & \frac{-g\mathcal{B}\sin\delta}{2}  & \frac{m^2}{2\omega}
	                                \end{array} \right)\Bigg] \left(\!\! \begin{array}{c}A_{1}(z) \\ A_{2}(z) \\\phi(z)\end{array}  \!\!\right) = 0,\label{eq:eom_genplanewaves}
	    \ee
	We introduce $a_1(z)$, $a_2(z)$, and $\chi(z)$ such that we remove the $e^{i\omega z}$-dependence of the solutions:
	\be
	\left(\!\! \begin{array}{c}A_{1}(z) \\ A_{2}(z) \\\phi(z)\end{array}  \!\!\right) = \left(\!\! \begin{array}{c}a_{1}(z) \\ a_{2}(z) \\\ \chi(z)\end{array}  \!\!\right)e^{i\omega z},
	\ee
	and then rotate by $(\frac{\pi}{2}-\delta)$ to an appropriate basis $(\vec{e_{\perp}}, \vec{e_{\parallel}})$, such that $\vec{\mathcal{B}} = (0, \mathcal{B})$. Solving the equations in a way similar to the one used in Section~\ref{sec:pw}, and going back to the $(\vec{e_{1}}, \vec{e_{2}})$ basis, we finally obtain:
	\be
		\left(\!\! \begin{array}{c}a_{1}(z) \\ a_{2}(z) \\\ \chi(z)\end{array}  \!\!\right) =
		\left(
		\begin{array}{ccc}
		K_{11}	&K_{12}	&K_{13}\\
		K_{12}	&K_{22}	&K_{23}\\
		K_{13}	&K_{23}	&K_{33}
		\end{array} \right)
		\left(\!\! \begin{array}{c}a_{1}(0) \\ a_{2}(0) \\\ \chi(0)\end{array}  \!\!\right)
	\ee
	with
	\be
		\left\{
		\begin{array}{llllll}
		K_{11} = \sin^2\delta\ e^{-i\frac{{\omega_p}^2}{2\omega} z} + \cos^2\delta \left({(\textrm{c}_{mix})}^2\ e^{-i \frac{{\mu_{C}}^2}{2\omega}z} + {(\textrm{s}_{mix})}^2\ e^{-i \frac{{\mu_{D}}^2}{2\omega}z}\right)\\
		K_{22} = \cos^2\delta\ e^{-i\frac{{\omega_p}^2}{2\omega} z} + \sin^2\delta \left({(\textrm{c}_{mix})}^2\ e^{-i \frac{{\mu_{C}}^2}{2\omega}z} + {(\textrm{s}_{mix})}^2\ e^{-i \frac{{\mu_{D}}^2}{2\omega}z}\right)\\
		K_{12} = -\sin\delta\cos\delta\ e^{-i\frac{{\omega_p}^2}{2\omega} z} + \cos\delta\sin\delta \left({(\textrm{c}_{mix})}^2\ e^{-i \frac{{\mu_{C}}^2}{2\omega}z} + {(\textrm{s}_{mix})}^2\ e^{-i \frac{{\mu_{D}}^2}{2\omega}z}\right)\\
		K_{13} = \cos\delta\ \frac{\sin(2\theta_{mix})}{2}\left(e^{-i \frac{{\mu_{C}}^2}{2\omega}z} - e^{-i \frac{{\mu_{D}}^2}{2\omega}z}\right)\\
		K_{23} = \sin\delta\ \frac{\sin(2\theta_{mix})}{2}\left(e^{-i \frac{{\mu_{C}}^2}{2\omega}z} - e^{-i \frac{{\mu_{D}}^2}{2\omega}z}\right)\\
		K_{33} = {(\textrm{s}_{mix})}^2\ e^{-i \frac{{\mu_{C}}^2}{2\omega}z} + {(\textrm{c}_{mix})}^2\ e^{-i \frac{{\mu_{D}}^2}{2\omega}z},\\
		\end{array}
		\right.
	\ee
	where $\mu_C$ and $\mu_D$ are respectively $\mu_+$ and $\mu_-$ when $\omega_p>m$, and the other way around when $m>\omega_p$.

	When we consider light traveling through regions of different magnetic fields, we use this result inside each region, ensuring the continuity of the fields at the boundaries and neglecting reflected waves, which have an amplitude of order $\left(\frac{r_{mix}}{\omega^2}\right)$.

	Note that for the `patchy' model, the magnetic field from cell to cell is not only rotated in the transverse plane, but can undergo the most general tridimensional rotation. As inside each region only the total transverse field is relevant, this gives lower transverse field strengths. When we have allowed an additional underlying field, we have kept it in the $\vec{e}_2$ direction throughout and calculated the angle $\delta$ for each region.

\bibliographystyle{mod}
\bibliography{alexbib}

\providecommand{\href}[2]{#2}\begingroup\raggedright\begin{thebibliography}{10}

\bibitem{Kim:1979}
J.~E. Kim, {\it {Weak Interaction Singlet and Strong $CP$ Invariance}},  {\em
  Phys. Rev. Lett.} {\bf 43} (1979) 103.

\bibitem{SVZ:1979}
M.~A. Shifman, A.~I. Vainshtein, and V.~I. Zakharov, {\it {Can Confinement
  Ensure Natural $CP$ Invariance of Strong Interactions?}},  {\em Nucl. Phys.}
  {\bf B166} (1980) 493.

\bibitem{Zhitnitsky:1980}
A.~R. Zhitnitsky, {\it {On Possible Suppression of the Axion Hadron
  Interactions. (In Russian)}},  {\em Sov. J. Nucl. Phys.} {\bf 31} (1980) 260.

\bibitem{DFS:1981}
M.~Dine, W.~Fischler, and M.~Srednicki, {\it {A Simple Solution to the Strong
  $CP$ Problem with a Harmless Axion}},  {\em Phys. Lett.} {\bf B104} (1981)
  199.

\bibitem{Weinberg:1978}
S.~Weinberg, {\it {A New Light Boson?}},  {\em Phys. Rev. Lett.} {\bf 40}
  (1978) 223--226.

\bibitem{Wilczek:1978}
F.~Wilczek, {\it {Problem of Strong $P$ and $T$ Invariance in the Presence of
  Instantons}},  {\em Phys. Rev. Lett.} {\bf 40} (1978) 279--282.

\bibitem{PecceiQuinn:1977}
R.~D. Peccei and H.~R. Quinn, {\it {$CP$ Conservation in the Presence of
  Instantons}},  {\em Phys. Rev. Lett.} {\bf 38} (1977) 1440--1443.

\bibitem{Raffelt:2006rj}
G.~G. Raffelt, {\it {Axions: Motivation, limits and searches}},  {\em J. Phys.}
  {\bf A40} (2007) 6607--6620,
  [\href{http://xxx.lanl.gov/abs/hep-ph/0611118}{{\tt hep-ph/0611118}}].

\bibitem{Jaeckel:2010ni}
J.~Jaeckel and A.~Ringwald, {\it The low-energy frontier of particle physics},
  {\em Ann. Rev. Nucl. Part. Sci.} {\bf 60} (2010) 405--437,
  [\href{http://xxx.lanl.gov/abs/1002.0329}{{\tt arXiv:1002.0329}}].

\bibitem{Sikivie:1983ip}
P.~Sikivie, {\it {Experimental tests of the *invisible* axion}},  {\em Phys.
  Rev. Lett.} {\bf 51} (1983) 1415.

\bibitem{Raffelt:1987im}
G.~Raffelt and L.~Stodolsky, {\it {Mixing of the Photon with Low Mass
  Particles}},  {\em Phys. Rev.} {\bf D37} (1988) 1237.

\bibitem{DeAngelis:2007dy}
A.~De~Angelis, O.~Mansutti, and M.~Roncadelli, {\it {Evidence for a new light
  spin-zero boson from cosmological gamma-ray propagation?}},  {\em Phys. Rev.}
  {\bf D76} (2007) 121301, [\href{http://xxx.lanl.gov/abs/0707.4312}{{\tt
  arXiv:0707.4312}}].

\bibitem{Burrage:2009mj}
C.~Burrage, A.-C. Davis, and D.~J. Shaw, {\it {Active Galactic Nuclei Shed
  Light on Axion-like- Particles}},  {\em Phys. Rev. Lett.} {\bf 102} (2009)
  201101, [\href{http://xxx.lanl.gov/abs/0902.2320}{{\tt arXiv:0902.2320}}].

\bibitem{Fairbairn:2009zi}
M.~Fairbairn, T.~Rashba, and S.~V. Troitsky, {\it {Photon-axion mixing in the
  Milky Way and ultra-high-energy cosmic rays from BL Lac type objects -
  Shining light through the Universe}},
  \href{http://xxx.lanl.gov/abs/0901.4085}{{\tt arXiv:0901.4085}}.

\bibitem{Hutsemekers:1998}
D.~Hutsem{\'e}kers, {\it {Evidence for very large-scale coherent orientations
  of quasar polarization vectors}},  {\em Astron. Astrophys.} {\bf 332} (1998)
  410--428.

\bibitem{Hutsemekers:2001}
D.~Hutsem{\'e}kers and H.~Lamy, {\it {Confirmation of the existence of coherent
  orientations of quasar polarization vectors on cosmological scales}},  {\em
  Astron. Astrophys.} {\bf 367} (2001) 381--387,
  [\href{http://xxx.lanl.gov/abs/astro-ph/0012182}{{\tt astro-ph/0012182}}].

\bibitem{Jain:2003sg}
P.~Jain, G.~Narain, and S.~Sarala, {\it {Large Scale Alignment of Optical
  Polarizations from Distant QSOs using Coordinate Invariant Statistics}},
  {\em Mon. Not. Roy. Astron. Soc.} {\bf 347} (2004) 394,
  [\href{http://xxx.lanl.gov/abs/astro-ph/0301530}{{\tt astro-ph/0301530}}].

\bibitem{Hutsemekers:2005}
D.~Hutsem{\'e}kers, R.~Cabanac, H.~Lamy, and D.~Sluse, {\it {Mapping
  extreme-scale alignments of quasar polarization vectors}},  {\em Astron.
  Astrophys.} {\bf 441} (2005) 915--930,
  [\href{http://xxx.lanl.gov/abs/astro-ph/0507274}{{\tt astro-ph/0507274}}].

\bibitem{Rusk:1985}
R.~Rusk and E.~Seaquist, {\it {Alignment of radio and optical polarization with
  VLBI structure}},  {\em Astron. J.} {\bf 90} (1985) 30--38.

\bibitem{Joshi:2007yf}
S.~A. Joshi {\em et~al.}, {\it {A survey of polarization in the JVAS/CLASS
  flat-spectrum radio source surveys -- II. A search for aligned radio
  polarizations}},  {\em Mon. Not. Roy. Astron. Soc.} {\bf 380} (2007)
  162--174, [\href{http://xxx.lanl.gov/abs/0705.2548}{{\tt arXiv:0705.2548}}].

\bibitem{Borguet:2007kb}
B.~Borguet, D.~Hutsem{\'e}kers, G.~Letawe, Y.~Letawe, and P.~Magain, {\it
  {Evidence for a Type 1/Type 2 dichotomy in the correlation between quasar
  optical polarization and host galaxy/extended emission position angles}},
  {\em Astron. Astrophys.} {\bf 478} (2008) 321--333. And references therein.

\bibitem{Borguet:2008tn}
B.~Borguet, D.~Hutsem{\'e}kers, G.~Letawe, Y.~Letawe, and P.~Magain, {\it {New
  Insights into the Quasar Type1/Type2 Dichotomy from Correlations between
  Quasar Host Orientation and Polarization}},  in {\em Astronomical Polarimetry
  2008: Science from Small to Large Telescopes} (P.~Bastien and N.~Manset,
  eds.), ASP Conf. Series, 2008.
\newblock \href{http://xxx.lanl.gov/abs/0809.4539}{{\tt arXiv:0809.4539}}.
\newblock To be published.

\bibitem{Jain:2002vx}
P.~Jain, S.~Panda, and S.~Sarala, {\it {Electromagnetic polarization effects
  due to axion photon mixing}},  {\em Phys. Rev.} {\bf D66} (2002) 085007,
  [\href{http://xxx.lanl.gov/abs/hep-ph/0206046}{{\tt hep-ph/0206046}}].

\bibitem{Das:2004qka}
S.~Das, P.~Jain, J.~P. Ralston, and R.~Saha, {\it {Probing dark energy with
  light: Propagation and spontaneous polarization}},  {\em JCAP} {\bf 0506}
  (2005) 002, [\href{http://xxx.lanl.gov/abs/hep-ph/0408198}{{\tt
  hep-ph/0408198}}].

\bibitem{Piotrovich:2008iy}
M.~Y. Piotrovich, Y.~N. Gnedin, and T.~M. Natsvlishvili, {\it
  {Photon-Axion-Like Particle Coupling Constant and Cosmological
  Observations}},  \href{http://xxx.lanl.gov/abs/0805.3649}{{\tt
  arXiv:0805.3649}}.

\bibitem{Payez:2008pm}
A.~Payez, J.~R. Cudell, and D.~Hutsem{\'e}kers, {\it {Axions and polarisation
  of quasars}},  {\em AIP Conf. Proc.} {\bf 1038} (2008) 211--219,
  [\href{http://xxx.lanl.gov/abs/0805.3946}{{\tt arXiv:0805.3946}}]. A
  discussion on the implications of dichroism and birefringence is found in
  sections 3 and 4.

\bibitem{Hutsemekers:2008iv}
D.~Hutsem{\'e}kers, A.~Payez, R.~Cabanac, H.~Lamy, D.~Sluse, B.~Borguet, and
  J.~R. Cudell, {\it {Large-Scale Alignments of Quasar Polarization Vectors:
  Evidence at Cosmological Scales for Very Light Pseudoscalar Particles Mixing
  with Photons?}},  in {\em Astronomical Polarimetry 2008: Science from Small
  to Large Telescopes} (P.~Bastien and N.~Manset, eds.), ASP Conf. Series,
  2008.
\newblock \href{http://xxx.lanl.gov/abs/0809.3088}{{\tt arXiv:0809.3088}}.
\newblock To be published.

\bibitem{Payez:2009kc}
A.~Payez, J.~R. Cudell, and D.~Hutsem{\'e}kers, {\it {Axion-like particles and
  circular polarisation of active galactic nuclei}},  in {\em Proceedings of
  the 5th Patras Workshop on Axions, WIMPs and WISPs} (J.~Jaeckel, A.~Lindner,
  and J.~Redondo, eds.), Verlag Deutsches Elektronen-Synchrotron, 2009.
\newblock \href{http://xxx.lanl.gov/abs/0911.0585}{{\tt arXiv:0911.0585}}.

\bibitem{Payez:2009vi}
A.~Payez, J.~R. Cudell, and D.~Hutsem{\'e}kers, {\it {On the circular
  polarisation of light from axion-photon mixing}},  {\em AIP Conf. Proc.} {\bf
  1241} (2010) 444--449, [\href{http://xxx.lanl.gov/abs/0911.3145}{{\tt
  arXiv:0911.3145}}].

\bibitem{Payez:2010xb}
A.~Payez, D.~Hutsem{\'e}kers, and J.~R. Cudell, {\it {Large-scale coherent
  orientations of quasar polarisation vectors: interpretation in terms of
  axion-like particles}},  {\em AIP Conf. Proc.} {\bf 1274} (2010) 144--149,
  [\href{http://xxx.lanl.gov/abs/1003.2213}{{\tt arXiv:1003.2213}}].

\bibitem{Agarwal:2009ic}
N.~Agarwal, A.~Kamal, and P.~Jain, {\it {Alignments in quasar polarizations:
  pseudoscalar-photon mixing in the presence of correlated magnetic fields}},
  {\em Phys. Rev.} {\bf D83} (2011) 065014,
  [\href{http://xxx.lanl.gov/abs/0911.0429}{{\tt arXiv:0911.0429}}].

\bibitem{Hutsemekers:2010fw}
D.~Hutsem{\'e}kers, B.~Borguet, D.~Sluse, R.~Cabanac, and H.~Lamy, {\it Optical
  circular polarization in quasars},  {\em Astron. Astrophys.} {\bf 520} (2010)
  L7, [\href{http://xxx.lanl.gov/abs/1009.4049}{{\tt arXiv:1009.4049}}].

\bibitem{Biggio:2006im}
C.~Biggio, E.~Masso, and J.~Redondo, {\it {Mixing of photons with massive
  spin-two particles in a magnetic field}},  {\em Phys. Rev.} {\bf D79} (2009)
  015012, [\href{http://xxx.lanl.gov/abs/hep-ph/0604062}{{\tt
  hep-ph/0604062}}]. See discussion in section 2.

\bibitem{Anderson:1963pc}
P.~W. Anderson, {\it {Plasmons, Gauge Invariance, and Mass}},  {\em Phys. Rev.}
  {\bf 130} (1963) 439--442.

\bibitem{Carlson:1994yqa}
E.~D. Carlson and W.~D. Garretson, {\it {Photon to pseudoscalar conversion in
  the interstellar medium}},  {\em Phys. Lett.} {\bf B336} (1994) 431--438.

\bibitem{Deffayet:2001pc}
C.~Deffayet, D.~Harari, J.-P. Uzan, and M.~Zaldarriaga, {\it {Dimming of
  supernovae by photon - pseudoscalar conversion and the intergalactic
  plasma}},  {\em Phys. Rev.} {\bf D66} (2002) 043517,
  [\href{http://xxx.lanl.gov/abs/hep-ph/0112118}{{\tt hep-ph/0112118}}].

\bibitem{Vallee:1990}
J.~P. Vallee, {\it {A possible excess rotation measure and large-scale magnetic
  field in the Virgo supercluster of galaxies}},  {\em Astron. J.} {\bf 99}
  (1990) 459--462.

\bibitem{Vallee:2002}
J.~P. Vallee, {\it {Faraday screen and reversal of rotation measure in the
  local supercluster plane}},  {\em Astron. J.} {\bf 124} (2002) 1322--1327.

\bibitem{Giovannini:2003yn}
M.~Giovannini, {\it {The magnetized universe}},  {\em Int. J. Mod. Phys.} {\bf
  D13} (2004) 391--502, [\href{http://xxx.lanl.gov/abs/astro-ph/0312614}{{\tt
  astro-ph/0312614}}].

\bibitem{Burrage:2008ii}
C.~Burrage, A.-C. Davis, and D.~J. Shaw, {\it {Detecting Chameleons: The
  Astronomical Polarization Produced by Chameleon-like Scalar Fields}},  {\em
  Phys. Rev.} {\bf D79} (2009) 044028,
  [\href{http://xxx.lanl.gov/abs/0809.1763}{{\tt arXiv:0809.1763}}].

\bibitem{Vallee:2011}
J.~P. Vallee, {\it {Magnetic fields in the galactic Universe, as observed in
  supershells, galaxies, intergalactic and cosmic realms}},  {\em New Astron.
  Rev.} {\bf 55} (2011) 91--154.

\bibitem{Kravtsov:2002ac}
A.~V. Kravtsov, A.~A. Klypin, and Y.~Hoffman, {\it {Constrained Simulations of
  the Real Universe: II. Observational Signatures of Intergalactic Gas in the
  Local Supercluster Region}},  {\em Astron. J.} {\bf 571} (2002) 563--575,
  [\href{http://xxx.lanl.gov/abs/astro-ph/0109077}{{\tt astro-ph/0109077}}].

\bibitem{Stockman:1984qz}
H.~S. Stockman, R.~L. Moore, and J.~R.~P. Angel, {\it The optical polarization
  properties of ``normal'' quasars},  {\em Astroph. J.} {\bf 279} (1984)
  485--498.

\bibitem{Berriman:1990qz}
G.~Berriman, G.~D. Schmidt, S.~C. West, and H.~S. Stockman, {\it An optical
  polarization survey of the palomar-green bright quasar sample},  {\em
  Astrophys. J. Suppl. Ser.} {\bf 74} (1990) 869--883.

\bibitem{Sluse:2005}
D.~Sluse, D.~Hutsem{\'e}kers, H.~Lamy, R.~Cabanac, and H.~Quintana, {\it {New
  optical polarization measurements of quasi-stellar objects. The data}},  {\em
  Astron. Astrophys.} {\bf 433} (2005) 757--764,
  [\href{http://xxx.lanl.gov/abs/astro-ph/0507023}{{\tt astro-ph/0507023}}].

\bibitem{Csaki:2001jk}
C.~Csaki, N.~Kaloper, and J.~Terning, {\it {Effects of the intergalactic plasma
  on supernova dimming via photon axion oscillations}},  {\em Phys. Lett.} {\bf
  B535} (2002) 33--36, [\href{http://xxx.lanl.gov/abs/hep-ph/0112212}{{\tt
  hep-ph/0112212}}].

\bibitem{Mortsell:2002dd}
E.~Mortsell, L.~Bergstrom, and A.~Goobar, {\it {Photon axion oscillations and
  type Ia supernovae}},  {\em Phys. Rev.} {\bf D66} (2002) 047702,
  [\href{http://xxx.lanl.gov/abs/astro-ph/0202153}{{\tt astro-ph/0202153}}].

\bibitem{DeAngelis:2007yu}
A.~De~Angelis, O.~Mansutti, and M.~Roncadelli, {\it {Axion-Like Particles,
  Cosmic Magnetic Fields and Gamma-Ray Astrophysics}},  {\em Phys. Lett.} {\bf
  B659} (2008) 847--855, [\href{http://xxx.lanl.gov/abs/0707.2695}{{\tt
  arXiv:0707.2695}}].

\bibitem{Mirizzi:2006zy}
A.~Mirizzi, G.~G. Raffelt, and P.~D. Serpico, {\it {Photon axion conversion in
  intergalactic magnetic fields and cosmological consequences}},  {\em Lect.
  Notes Phys.} {\bf 741} (2008) 115--134,
  [\href{http://xxx.lanl.gov/abs/astro-ph/0607415}{{\tt astro-ph/0607415}}].

\bibitem{Hutsemekers:1998pp}
D.~Hutsem{\'e}kers, H.~Lamy, and M.~Remy, {\it {Polarization properties of a
  sample of broad absorption line and gravitationally lensed quasars}},  {\em
  Astron. Astrophys.} {\bf 340} (1998) 371--380.

\bibitem{Schmidt:1999p}
G.~Schmidt and D.~Hines, {\it {The Polarization of Broad Absorption Line
  QSOs}},  {\em Astron. J.} {\bf 512} (1999) 125--135.

\bibitem{Lamy:2004yz}
H.~Lamy and D.~Hutsem{\'e}kers, {\it {Polarization properties of broad
  absorption line QSOs : new statistical clues}},  {\em Astron. Astrophys.}
  {\bf 427} (2004) 107--123,
  [\href{http://xxx.lanl.gov/abs/astro-ph/0408476}{{\tt astro-ph/0408476}}].

\bibitem{Maiani:1986md}
L.~Maiani, R.~Petronzio, and E.~Zavattini, {\it {Effects of nearly massless,
  spin zero particles on light propagation in a magnetic field}},  {\em Phys.
  Lett.} {\bf B175} (1986) 359.

\bibitem{Cameron:1993mr}
R.~Cameron {\em et~al.}, {\it {Search for nearly massless, weakly coupled
  particles by optical techniques}},  {\em Phys. Rev.} {\bf D47} (1993)
  3707--3725.

\bibitem{Sterken:1992ap}
C.~Sterken and J.~Mandroid, {\em Astronomical Photometry --- A Guide}.
\newblock Kluwer Academic, 1992.

\bibitem{EFOSC2:2008}
L.~Monaco and C.~Snodgrass, {\em {EFOSC2 User's Manual}}.
\newblock European Southern Observatory, 2008.
\newblock LSO-MAN-ESO-36100-0004.

\bibitem{Shakura:1973bh}
N.~I. Shakura and R.~A. Sunyaev, {\it {Black Holes in Binary Systems.
  Observational Appearance.}},  {\em Astron. Astrophys.} {\bf 24} (1973)
  337--355.

\bibitem{Kochanek:2007ta}
C.~S. Kochanek, X.~Dai, C.~Morgan, N.~Morgan, S.~Poindexter, and G.~Chartas,
  {\it {Turning AGN microlensing from a curiosity into a tool}},  in {\em
  Statistical Challenges in Modern Astronomy IV} (G.~J. Babu and E.~D.
  Feigelson, eds.), vol.~371 of {\em ASP Conf. Series}.
\newblock \href{http://xxx.lanl.gov/abs/astro-ph/0609112}{{\tt
  astro-ph/0609112}}.

\bibitem{Donges:1998}
A.~Donges, {\it The coherence length of black-body radiation},  {\em Eur. J.
  Phys} {\bf 19} (1998) 245--249.

\bibitem{mpfr}
L.~Frousse, G.~Hanrot, V.~Lefèvre, P.~Pélissier, and P.~Zimmermann, {\it
  {MPFR: A multiple-precision binary floating-point library with correct
  rounding}},  {\em ACM Trans. Math. Softw.} {\bf 33} (2007) 15.

\bibitem{Landstreet:1972so}
J.~D. Landstreet and J.~R.~P. Angel, {\it {Search for optical circular
  polarization in quasars and Seyfert nuclei}},  {\em Astroph. J.} {\bf 174}
  (1972) L127--L129.

\bibitem{Stodolsky:1998tc}
L.~Stodolsky, {\it When the wavepacket is unnecessary},  {\em Phys. Rev.} {\bf
  D58} (1998) 036006, [\href{http://xxx.lanl.gov/abs/hep-ph/9802387}{{\tt
  hep-ph/9802387}}].

\bibitem{Bassan:2010ya}
N.~Bassan, A.~Mirizzi, and M.~Roncadelli, {\it {Axion-like particle effects on
  the polarization of cosmic high-energy gamma sources}},  {\em JCAP} {\bf
  1005} (2010) 010, [\href{http://xxx.lanl.gov/abs/1001.5267}{{\tt
  arXiv:1001.5267}}].

\bibitem{Serkowski:1958sa}
K.~Serkowski, {\it Statistical analysis of the polarization and reddening of
  the double cluster in perseus},  {\em Acta Astron.} {\bf 8} (1958) 135--170.
  See Appendix I.

\bibitem{Dolag:2003ft}
K.~Dolag, D.~Grasso, V.~Springel, and I.~Tkachev, {\it {Constrained simulations
  of the magnetic field in the local supercluster and the propagation of
  UHECR}},  in {\em 28th International Cosmic Ray Conference}, pp.~735--738,
  ICRR, University of Tokyo, 2003.
\newblock \href{http://xxx.lanl.gov/abs/astro-ph/0308155}{{\tt
  astro-ph/0308155}}.

\bibitem{Dolag:2005a}
K.~Dolag, D.~Grasso, V.~Springel, and I.~Tkachev, {\it {Simulating the magnetic
  field in the local supercluster}},  in {\em X-Ray and Radio Connections}
  (L.~Sjouwerman and K.~Dyer, eds.), 2004.
\newblock Published electronically at http://www.aoc.nrao.edu/events/xraydio.

\bibitem{tocome}
A.~Payez {\em et~al.} To be published.

\end{thebibliography}\endgroup

\end{document}